\documentclass[a4paper,11pt]{article}
\pdfoutput=1
\usepackage{jheppub}
\usepackage{textcase}
\usepackage{graphicx,epsfig,psfrag,amssymb,hyperref}
\usepackage{multirow}
\usepackage{color,graphicx,epsfig,psfrag,amsmath,empheq}
\usepackage[dvipsnames]{xcolor}
\usepackage{bm}
\usepackage{mathrsfs,amsfonts,soul,color}
\usepackage[caption=false]{subfig}
\usepackage{hepunits}
\usepackage{enumitem}
\usepackage{placeins}
\usepackage{textcomp}
\usepackage{comment}

\usepackage{float}
\usepackage[toc,page]{appendix}
\usepackage[ruled,vlined]{algorithm2e}
\usepackage{wrapfig}
\SetKwInOut{Input}{Input}
\SetKwInOut{Output}{Output}
\SetKwFunction{Fforward}{Forward}
\SetKwFunction{Fbackward}{Backward}
\SetKwProg{Fn}{function}{:}{}
\DontPrintSemicolon

\begin{document}

\title{Quantum Anomaly Detection for Collider Physics} 

\author{Sulaiman Alvi,$^{a,b}$}
\author{Christian Bauer,$^{b}$}
\author{and Benjamin Nachman,$^{b,c}$}

\affiliation{
\phantom{ }\hspace{-0.12in}$^b$Department of Physics, University of California, Berkeley, CA 94720, USA \\
\phantom{ }\hspace{-0.12in}$^b$Physics Division, Lawrence Berkeley National Laboratory, Berkeley, CA 94720, USA \\
\phantom{ }\hspace{-0.12in}$^c$Berkeley Institute for Data Science, University of California, Berkeley, CA 94720, USA
}

\emailAdd{sulaimanalvi@berkeley.edu}
\emailAdd{cwbauer@lbl.gov}
\emailAdd{bpnachman@lbl.gov}

\abstract{
Quantum Machine Learning (QML) is an exciting tool that has received significant recent attention due in part to advances in quantum computing hardware.  While there is currently no formal guarantee that QML is superior to classical ML for relevant problems, there have been many claims of an empirical advantage with high energy physics datasets.  These studies typically do not claim an exponential speedup in training, but instead usually focus on an improved performance with limited training data.  We explore an analysis that is characterized by a low statistics dataset.  In particular, we study an anomaly detection task in the four-lepton final state at the Large Hadron Collider that is limited by a small dataset.  We explore the application of QML in a semi-supervised mode to look for new physics without specifying a particular signal model hypothesis.  We find no evidence that QML provides any advantage over classical ML.  It could be that a case where QML is superior to classical ML for collider physics will be established in the future, but for now, classical ML is a powerful tool that will continue to expand the science of the LHC and beyond.
}

\maketitle

%%%%%%%%%%%%%%%%%%%%%%%%%%%%%%%%%%%%%%%%
\section{Introduction}
\label{sec:intro}
%%%%%%%%%%%%%%%%%%%%%%%%%%%%%%%%%%%%%%%%

Quantum computing is a promising tool for a variety of problems because an exponentially large Hilbert space can be described by polynomially many qubits.  In high energy physics, there is particular promise for simulations of quantum field theories, where every spacetime point has quantum degrees of freedom, but polynomial algorithms exist for state preparation and time evolution~\cite{Jordan:2012xnu,Bauer:2022hpo}.  However, not all classically hard algorithms are more efficient on a quantum computer.

One particular class of algorithms that has received significant attention in high energy physics (HEP) is Quantum Machine Learning (QML).  In this paper, QML refers to machine learning tasks that are executed on quantum computing hardware.  While QML is not known to be more efficient than classical machine learning (CML), there have been many empirical studies to explore the potential of QML for HEP~\cite{Mott:2017xdb,Zlokapa:2019lvv,Blance:2020nhl,Terashi:2020wfi,Chen:2020zkj,Wu:2020cye,Chen:2021ouz,Heredge:2021vww,Wu:2021xsj,Belis:2021zqi,Araz:2021ifk,Bravo-Prieto:2021ehz,Blance:2021gcs,Ngairangbam:2021yma,Araz:2022haf,Gianelle:2022unu,Humble:2022vtm} (see also Ref.~\cite{Guan:2020bdl} for a recent review).

A common theme that has emerged from these studies is that QML seems to outperform CML with small training datasets.  While there is no rigorous explanation for this observation, it could be that QML provides a superior inductive bias and/or more expresivity with a smaller number of parameters.  In nearly all studies, CML outperforms QML when there are more than $\mathcal{O}(100)$ examples.  There are almost no problems in collider HEP that have such small numbers of events for training.  The goal of this paper is to explore a realistic use case for near-term QML for collider physics. See also Ref.~\cite{Schuld:2022lss} for the broader context of QML versus CML.

Most analyses at the Large Hadron Collider (LHC) make use of simulation for training classifiers.  Since these simulations can be used to generate more events independent of collider operations, they are not in a regime where QML is expected to outperform CML with near-term quantum hardware.  Therefore, analyses that have the potential for QML to outperform CML should require data for training.  Since we do not know the origin of any particular data event, such methods are called \textit{less than supervised}~\cite{Karagiorgi:2021ngt}.  

One particularly promising class of less-than-supervised methods is signal model-independent anomaly detection.  Such approaches are characterized by the comparison of data in a particular region of phase space (\textit{signal region}) with a reference sample. There have been many proposals for doing this comparison using machine learning~\cite{Collins:2018epr,Collins:2019jip,DAgnolo:2018cun,DAgnolo:2019vbw,Andreassen:2020nkr,Nachman:2020lpy,Hallin:2021wme,Farina:2018fyg,Heimel:2018mkt,Roy:2019jae,Cerri:2018anq,Blance:2019ibf,Hajer:2018kqm,DeSimone:2018efk,Mullin:2019mmh,1809.02977,Dillon:2019cqt,Aguilar-Saavedra:2017rzt,Romao:2019dvs,Romao:2020ojy,knapp2020adversarially,collaboration2020dijet,1797846,1800445,Amram:2020ykb,Cheng:2020dal,Khosa:2020qrz,Thaprasop:2020mzp,Alexander:2020mbx,aguilarsaavedra2020mass,Aguilar-Saavedra:2021utu,1815227,pol2020anomaly,Mikuni:2020qds,vanBeekveld:2020txa,Park:2020pak,Faroughy:2020gas,Stein:2020rou,Kasieczka:2021xcg,Batson:2021agz,Blance:2021gcs,Bortolato:2021zic,Collins:2021nxn,Dillon:2021nxw,Finke:2021sdf,Shih:2021kbt,Atkinson:2021nlt,Kahn:2021drv,Aarrestad:2021oeb,Chakravarti:2021svb,Dorigo:2021iyy,Caron:2021wmq,Govorkova:2021hqu,Kasieczka:2021tew,dAgnolo:2021aun,Volkovich:2021txe,Govorkova:2021utb,Ostdiek:2021bem,Fraser:2021lxm,Raine:2022hht,Krzyzanska:2022mto,Letizia:2022xbe,Letizia2021}.  A highly sensitive approach is to train a classifier to distinguish data from a precise prediction of the background (semi-supervised learning)~\cite{Collins:2018epr,Collins:2019jip,DAgnolo:2018cun,DAgnolo:2019vbw,Andreassen:2020nkr,dAgnolo:2021aun,Chakravarti:2021svb,Nachman:2020lpy,Hallin:2021wme,Krzyzanska:2022mto,Letizia:2022xbe,Letizia2021,Raine:2022hht}.  If the background is well-understood theoretically, then the reference sample could be simulation~\cite{DAgnolo:2018cun,DAgnolo:2019vbw,dAgnolo:2021aun,Chakravarti:2021svb,Krzyzanska:2022mto,Letizia:2022xbe,Letizia2021}. This has the advantage that the background prediction does not need to be learned, but has the disadvantage of being strongly background-model dependent.

There are few final states at the LHC for which the background is known precisely enough to be used directly for background estimation.  One exception is the final state with four charged leptons.  Both ATLAS~\cite{ATLAS:2020tlo,ATLAS:2018coo,ATLAS:2020wny} and CMS~\cite{CMS:2016ilx,CMS:2020bni,CMS:2021nnc} directly use Monte Carlo (MC) simulations to estimate the background and ATLAS even uses machine learning to isolate particular signals~\cite{ATLAS:2020tlo}.  While powerful, this approach is signal model-specific and does not readily extend to models with multidimensional parameters.  Reference~\cite{Krzyzanska:2022mto} recently proposed to use machine learning as an alternative approach.  It was shown that training classifiers to distinguish data from background-only simulation provides a complementary approach to direct searches and has broad signal sensitivity.  As the four lepton final state also has a small cross section, this is a natural target for studying QML.

The prospect of QML for anomaly detection was first studied in Ref.~\cite{Ngairangbam:2021yma} in the context of autoencoders.  These unsupervised tools can be trained without any simulation, but are not as effective as semi-supervised methods when there is a good background model and/or when the new physics is not the lowest density events~\cite{Collins:2021nxn,Fraser:2021lxm}.  For this reason, our focus is on semi-supervised learning.  We also assume an idealized situation where there are no systematic uncertainties.  Nuisance parameters will likely not change the qualitative conclusions of the QML versus CML comparison and have been discussed in Ref.~\cite{dAgnolo:2021aun} for anomaly detection. 

This paper is organized as follows.  Section~\ref{sec:qml} briefly introduces various QML approaches.  The simulated samples to be used for the machine learning are described in Sec.~\ref{sec:sim}.  Numerical results are presented in Sec.~\ref{sec:results} and the paper ends with conclusions and outlook in Sec.~\ref{sec:concl}.

\clearpage

%%%%%%%%%%%%%%%%%%%%%%%%%%%%%%%%%%%%%%%%
\section{Quantum Machine Learning}
\label{sec:qml}
%%%%%%%%%%%%%%%%%%%%%%%%%%%%%%%%%%%%%%%%

There are a number of ways quantum computers can be used for machine learning.  One possibility is that quantum algorithms can reduce the computational complexity of linear algebra operations core to a number of ML approaches (see e.g. Refs.~\cite{Rebentrost_2014,PhysRevLett.103.150502,Biamonte_2017}).  Another possibility is to use quantum circuits as flexible function approximators similar to neural networks and other classical ML techniques.  Quantum-classical variational methods used to optimize parameteric circuits are the analog of tuning the weights and biases of a classical neural network.  There have been a number of claims in the literature that QML methods of this kind outperform CML for limited training data.  While there is no formal proof of this claim, it could be justified intuitively as a result of the improved \textit{inductive bias} of quantum circuits.  In other words, the class of functions that QML represent with a limited set of parameters are more relevant / tailored to HEP problems.  We aim to explore this claim in the context of two variational algorithms in the four lepton anomaly detection search\footnote{As a cross-check, we also explored the Supersymmetry example of Ref.~\cite{Terashi:2020wfi}.  We were unable to reproduce the QML superior performance at low sample size and the CML was also significantly improved by adding more parameters.}.

The two QML methods we study are called Variational Quantum Circuits (VQC) and Quantum Circuit Learning (QCL).  These are both implementations of parameterized quantum circuits where the various rotation angles are optimized via classical methods.  Each algorithm is composed of multiple components:  state preparation, which encode classical data into quantum states, the model circuit, which contains the parameters that are optimized during the training process, and the measurement and output, which are used to evaluate performance of the circuit.  VQC and QCL differ only in the structure of the parameterized circuit, as detailed in Sec.~\ref{sec:mlsetup}.  As the examples we study in this paper are relatively small in terms of quantum resources, all circuits are simulated on classical computers.   

Note that we also studied Quantum Support Vector Machines (QSVMs)~\cite{PhysRevLett.103.150502}, but initial tests  suggested that they are strictly less effective than other methods so they were not included in the final tests.  See also Ref.~\cite{Paul-Aymeric} for a broader perspective on QVSMs. We have also explored the quantum gradient descent studied in Ref.~\cite{Blance:2020nhl}.  We tested the setup using \texttt{Pennylane} \cite{Pennylane} and found that the learning rates were unreliable for Gaussian classification as well as our HEP application, so this was not pursued further.

\clearpage

%%%%%%%%%%%%%%%%%%%%%%%%%%%%%%%%%%%%%%%%
\section{Simulation}
\label{sec:sim}
%%%%%%%%%%%%%%%%%%%%%%%%%%%%%%%%%%%%%%%%

The simulated datasets are the same as in Ref.~\cite{Krzyzanska:2022mto} and are briefly summarized in the following.  All events are generated with \textsc{MadGraph5\_aMC@NLO} 2.8.0~\cite{Alwall:2014hca}.  Both signal and background events are generated using the Higgs Boson Effective Field Theory (\texttt{heft}) in which the heavy top quark limit is used for the gluon-gluon-Higgs vertex.  We focus on the $e^+e^-\mu^+\mu^-$ final state to avoid combinatoric ambiguity.  After the matrix element calculations, the outgoing particles are processed with \textsc{Pythia}~8.244~\cite{Sjostrand:2006za,Sjostrand:2007gs,Sjostrand:2014zea} with its default settings for parton showering and hadronization.  \textsc{Pythia} also handles the decay of the anomalies, which are Higgs-like scalar particles with  a mass of 125 GeV decaying asymmetrically into two lighter-mass bosons, one of which decays to electrons and the other which decays into muons.  This is accomplished technically by generating Higgs bosons and then replacing the PDGID~\cite{Zyla:2020zbs} of 25 (SM Higgs) with 35 (2HDM heavier Higgs) in the Les Houches Event (LHE) files~\cite{Alwall:2006yp} and then setting the decay of this particle into particles with PDGIDs 23 (Z boson) and 36 (2HDM pseudoscalar).  Subsequently, the $Z$ boson (pseudoscalar) is forced to decay into electrons (muons).  All three BSM particles are set to a narrow width.  The detector response is emulated with \textsc{Delphes} 3.4.2~\cite{deFavereau:2013fsa,Mertens:2015kba,Selvaggi:2014mya} using the default CMS card.  We will make the simplifying assumption that there is no system uncertainty in the background estimation and so the `data' and `simulation' are statistically identical when no signal events are injected.  Recent studies exploring the integration of systematic uncertainties in this setup can be found in Ref.~\cite{dAgnolo:2021aun}.  The number of events in background corresponds to the LHC Run 2 dataset (about 150 fb$^{-1}$).

In this work we focus only on the leptoninc final states. Other event properties could also be useful, however, information about the hadronic final state is known with less precision and thus may introduce the need to go beyond a pure simulation-based background estimation. Each event is characterized four three-momenta (12 numbers in total).  For our target models, $pp\rightarrow A\rightarrow B(\rightarrow e^+e^-)C(\rightarrow \mu^+\mu^-)$, the three masses $m_{e^+e^-\mu^+\mu^-},m_{e^+e^-},m_{\mu^+\mu^-}$ are nearly sufficient statistics for characterizing the new physics.  In this paper, we consider signals of this form, where $A$ is a non-SM Higgs boson that decays to two different mass bosons.  The model is specified by three parameters: $m_A, m_B$, and $m_C$.  There is also an overall cross section set by the coupling of the $A$ particle to the rest of the Standard Model.  This cross section will be varied in the subsequent analysis by considering different numbers of signal events.  Given the three-dimensional parameter space, we focus on the three-dimensional problem in this paper.  Non-resonant signals and signals with non-trivial spin structures could benefit from using more of the phase space in the future.  While there are currently LHC searches for the case that $B=C$ (or $B$ is a BSM particle and $C$ is a $Z$ boson), there is currently on search where all three masses could be different.  This is not because physics motivation is lacking~\cite{Robens:2019kga}, but instead that even a three-dimensional traditional search strategy is too complicated.

\begin{figure}[h!]
    \centering
    \includegraphics[width=0.5\textwidth]{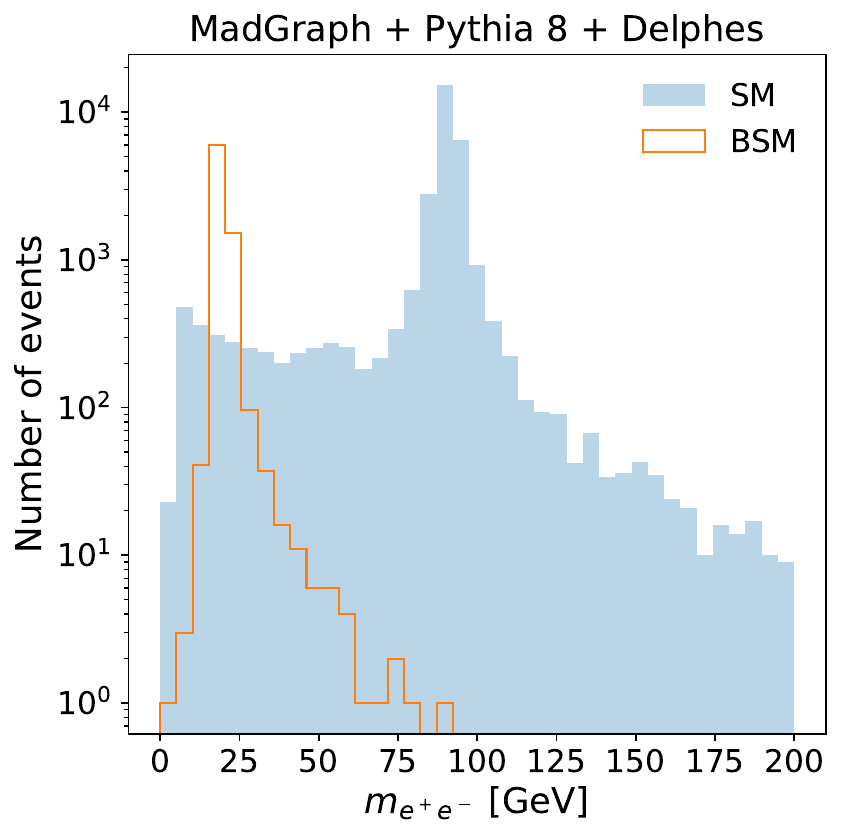}\includegraphics[width=0.5\textwidth]{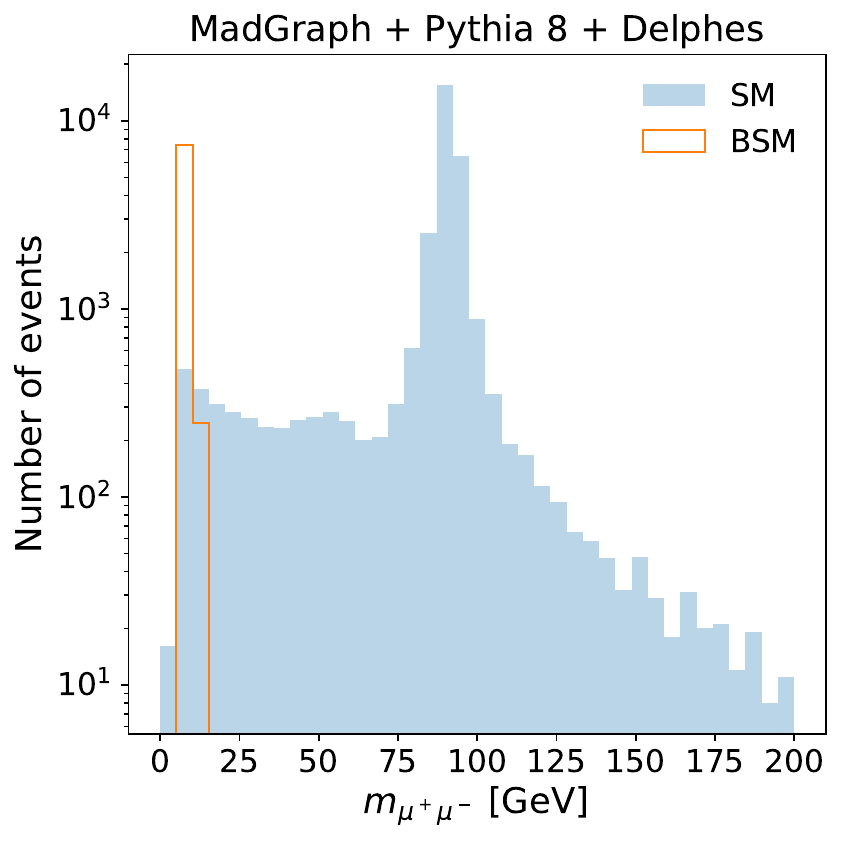}\\\includegraphics[width=0.5\textwidth]{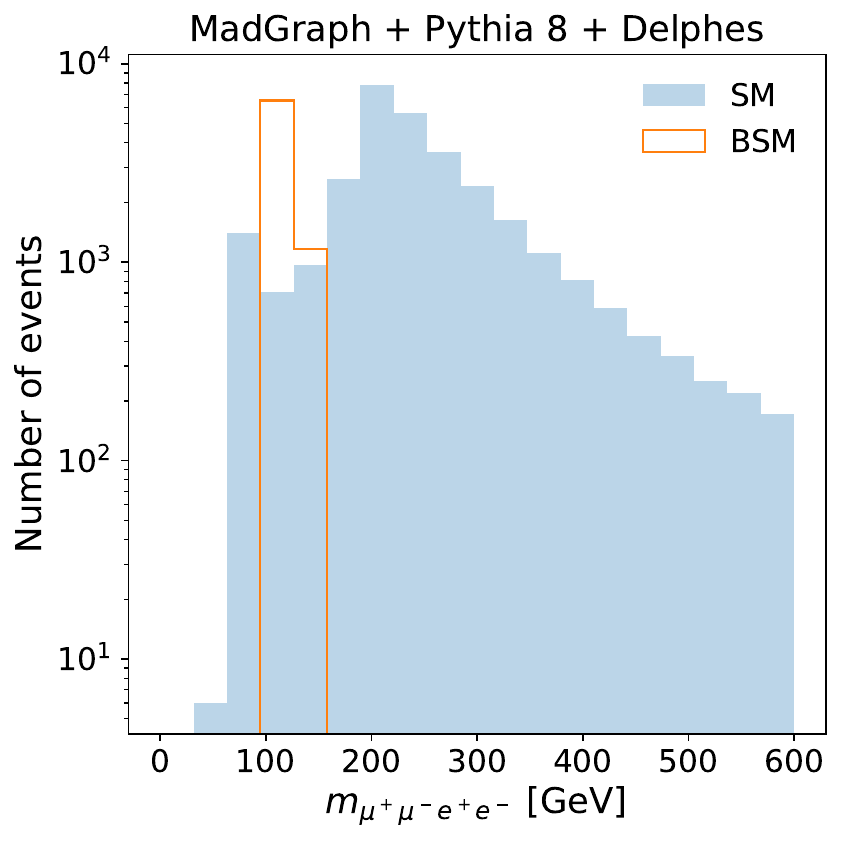}
    \caption{The three dimensions used for machine learning: $m_{e^+e^-}$ (left), $m_{\mu^+\mu^-}$ (right) and $m_{4\ell}$ (bottom).
    }
    \label{fig:inputs}
\end{figure}

The spectra of the three invariant masses ($m_{e^+e^-\mu^+\mu^-},m_{e^+e^-},m_{\mu^+\mu^-}$) for the background and our representative signal are presented in Fig.~\ref{fig:inputs}.  As expected, the di-electron and di-muon invariant masses peak near the $Z$ boson mass of 90 GeV~\cite{Zyla:2020zbs} and there are peaks in the four-lepton invariant mass at the $Z$ peak and the Higgs boson mass of about 125 GeV~\cite{Zyla:2020zbs}.  The signal is resonant in all three observables with peaks at the masses of the particles.  The parent particle (125 GeV mass) decays to two children, with masses of 25 and 15 GeV for electrons and muons, respectively. Note that these parameters are not known to the neural network.

\clearpage

\section{(Quantum) Machine Learning Setup}
\label{sec:mlsetup}

As stated earlier, our QML approaches are both variational circuits that are analogous to classical neural networks with a large number of tunable parameters.  We consider two flavors: VQC and QCL.  These two approaches are described in more detail below and only differ in how the classical data are encoded and what parameterized circuits are used in the learning.  VQC uses a simpler encoding with a multi-qubit rotation followed by a series of CNOT gates for the variational part.  Instead, QCL uses a more complex encoding and then time evolution of a certain Hamiltonian for the variational component. 

\subsection{VQC}

\begin{figure}[h!]
\centering
\includegraphics[width=0.85\textwidth]{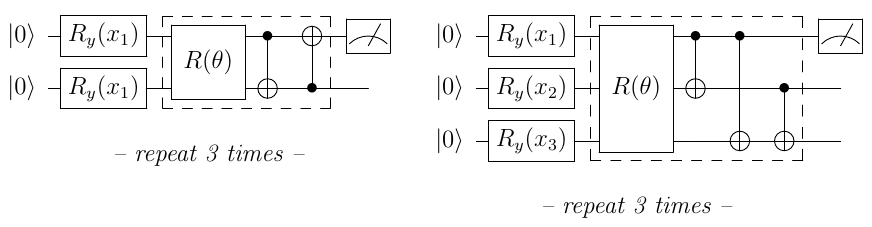}
\caption{VQC circuit for the one-dimensional (left) and three-dimensional (right) datasets.}
\label{fig:vqc1d}
\end{figure}

        Figure~\ref{fig:vqc1d} shows the VQC circuits used in this study.  The input features ($x_i$) are min-max scaled so that the argument of the initial $R_y$ gates are valid angles. The rotational gate $R(\theta)$ is given by $R_Z(\alpha)R_Y(\omega)R_Z(\phi)$, where $\alpha$, $\omega$, and $\phi$ are the trainable weights of the circuit.  These angles are unique for each qubit, leading to a total of 18 parameters for the one-dimensional setup and 27 for the three-dimensional setup.  The output is the expectation value of $Z$, which is achieved by taking several shots of each qubits. 
        
    There are a number of hyperparameters that must be selected and the ones described above were chosen after a detailed study.  By varying the number of qubits, we observed that two qubits resulted in an optimal accuracy for the one-dimensional classification task.  Using more than six qubits caused issues related to overfitting of the data while also increasing runtime.  As such we chose three qubits for the three-dimensional case.  We observed that we needed a minimum of three rotation gates in order to achieve an effective level of expressivity of the circuit.  Between three and five rotation gates, we did not see a significant increase in accuracy, and beyond five rotation gates, the time to run the simulation became prohibitive while also not resulting in an increase in performance.  The default encoding gate included a $Y$ and $Z$ rotation but testing different combinations of input encoding gates showed no performance difference so the final circuit just included a $Y$ rotation gate.
        Batch learning was also implemented for the three-dimensional dataset tests, which greatly improved performance. The circuit was optimized using vanilla gradient descent, with the quantum gradient calculated using \texttt{Pennylane}'s \cite{Pennylane} implementation of the parameter shift rules for quantum circuits.

\subsection{QCL}

\begin{figure}[h!]
\centering
\includegraphics[width=0.85\textwidth]{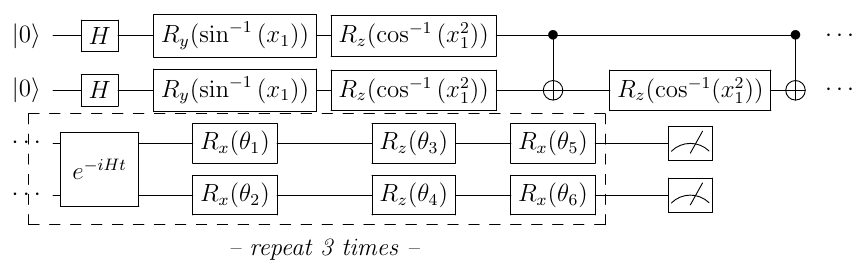}
\caption{QCL circuit for 1 dimensional datasets.}
\label{fig:qcl1d}
\end{figure}

The circuit's parameterized gates follow~\cite{Terashi:2020wfi} in a way that allows for access of the entire Bloch sphere given an arbitrary input state.  The gate that introduces entanglement involves a time evolution operation following the Ising model Hamiltonian $H$ as in Figure \ref{fig:qcl1d}.  We did not experiment with different Hamiltonians as their work remarks that changing this did not achieve a different result.  The expectation value $z_i$ of the Pauli Z operator for the first two qubits is measured and the loss function is cross-entropy:
        
        \begin{align}
            -\sum\left[ \log\left(\frac{e^{z_0}}{e^{z_0}+e^{z_1}}\right)+\log\left(\frac{e^{z_1}}{e^{z_0}+e^{z_1}}\right)\right],
        \end{align}
        where the exponential terms represent the soft-max activation function. %
The number of qubits to encode parameters in the 1D case was chosen after a brief exploration to be two.  Each input value was duplicated and stored in two separate qubits which were transformed by the same circuit.  For 3D testing we used six qubits.  For all tests the circuits had three layers, which are outlined in Figure \ref{fig:qcl1d}.  The optimization was completed using the \texttt{COBYLA} method~\cite{Powell1994} with the parameter shift rule as defined in ~\cite{Terashi:2020wfi} and binary cross-entropy loss. 

The $\theta_i$ are the trainable weights of the circuit.  The three-dimensional case follows the same structure as Figure 2 and the alternative $U_{\textbf{in}}$ in Figure 10 from~\cite{Terashi:2020wfi}, except the three inputs are duplicated so that six qubits can be used.  There were 18 trainable parameters for the one-dimensional setup and 54 trainable parameters for the three-dimensional setup.
            
\subsection{CML}
    
The performance of the quantum machine learning setups are compared against two neural networks implemented in \texttt{TensorFlow}~\cite{tensorflow2015-whitepaper}.  The first network (`NN Low') has one layer with 32 nodes and the second network (`NN High') had two 32 node layers, which corresponds to 97 and 1,217 trainable parameters, respectively.  The output was passed through a sigmoid activation and the NN High model used rectified linear activation (ReLU) functions for the intermediate layer.

The networks were optimized using the binary cross entropy with \texttt{Adam}~\cite{adam}. Each network was run for a fixed number of epochs, determined by the convergence of the loss curves: 50 epochs for the one-dimensional example and 100 for three-dimensional example.

\clearpage

%%%%%%%%%%%%%%%%%%%%%%%%%%%%%%%%%%%%%%%%
\section{Results}
\label{sec:results}

We perform a weakly/semi-supervised search where a classifier is trained to distinguish a background sample from another, statistically independent and identical sample, that has some number of signal events added to it.  Due to the small number of injected signal events, it is important to study the sensitivity to different random sets of signal events.  The results presented below are averaged over 80 different random selections of signal and background (with error bars representing the standard deviation).  The performance is quantified in terms of the number of standard deviations\footnote{This is approximated by the number of signal events divided by the square root of the number of background events.  We do not expect that more precise calculations/approximations will qualitatively change the results.} achieved after applying the optimal (maximum significance improvement) threshold on the network\footnote{Other papers have used the Area Under the Receiver Operator Characteristic Curve (ROC AUC).  While the AUC is a standard metric in machine learning, it is not representative for HEP applications where typically one working point (threshold) on the classifier is used instead of integrating across the entire ROC curve.}.  The maximum significance improvement is model dependent, but it is chosen here to bound the achievable performance.  In practice, the challenge of selecting the threshold is the same for both QML and CML methods.  Two relevant benchmark significance are $2\sigma$ and $5\sigma$ which approximately correspond to the community standards for excluding and discovering a model, respectively.

\begin{figure}[h!]
    \centering
    \includegraphics[width=0.5\textwidth]{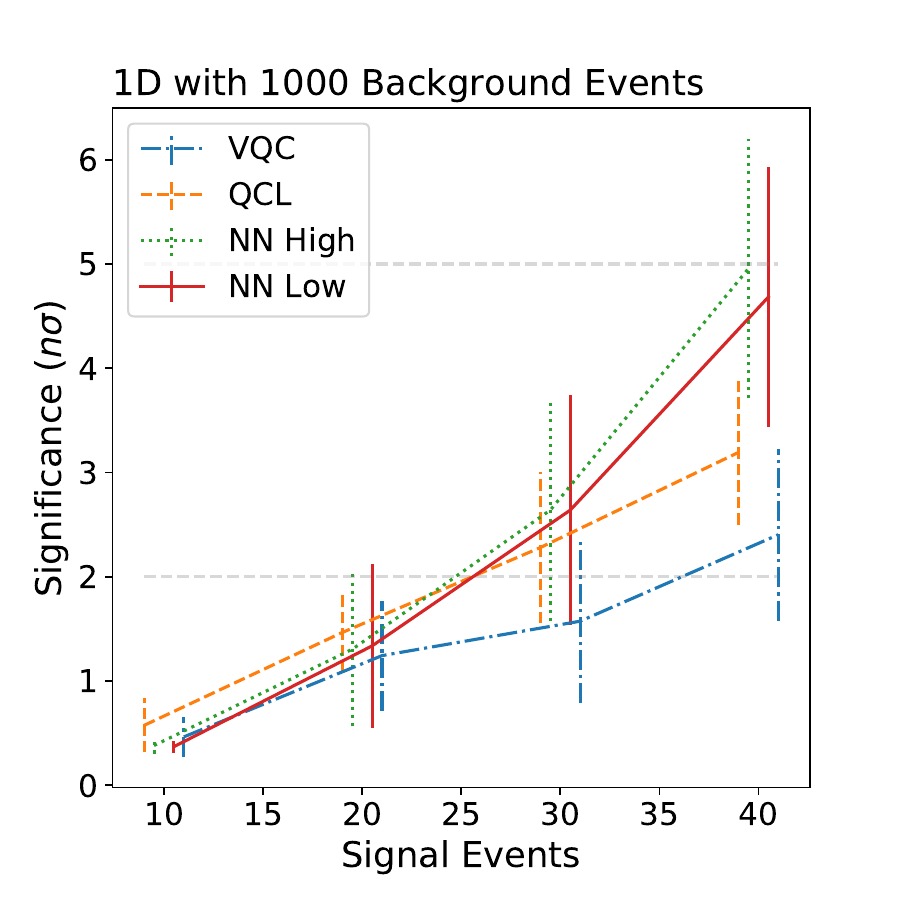}\includegraphics[width=0.5\textwidth]{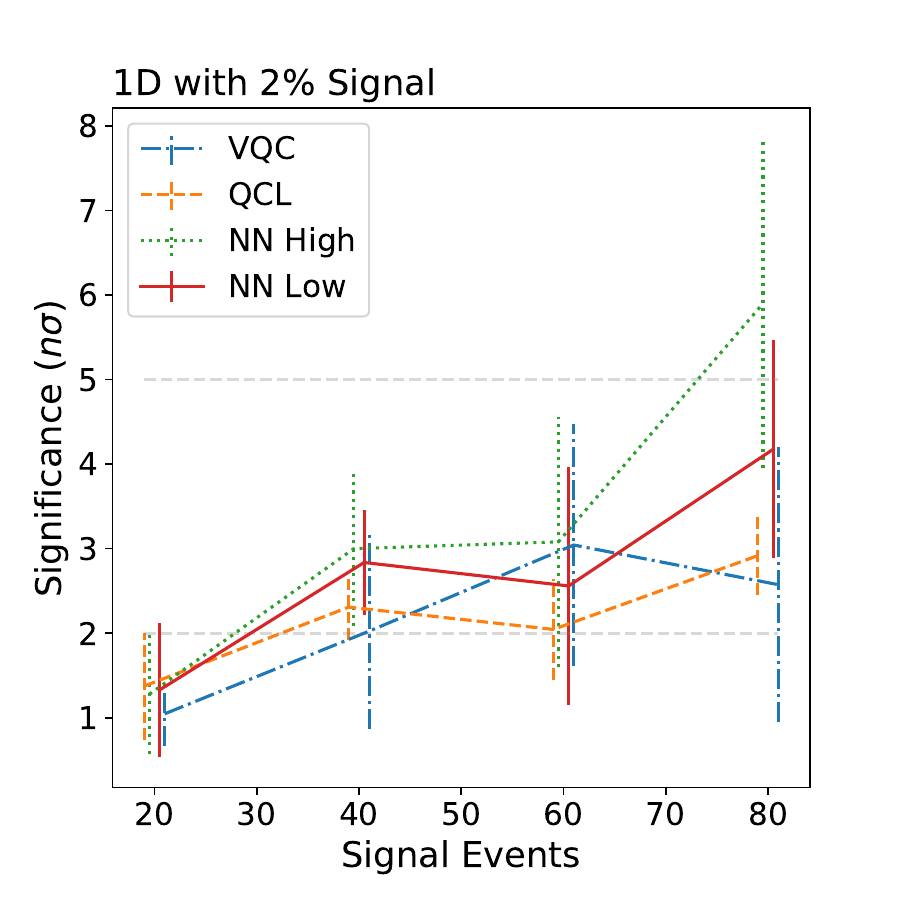}\\\includegraphics[width=0.5\textwidth]{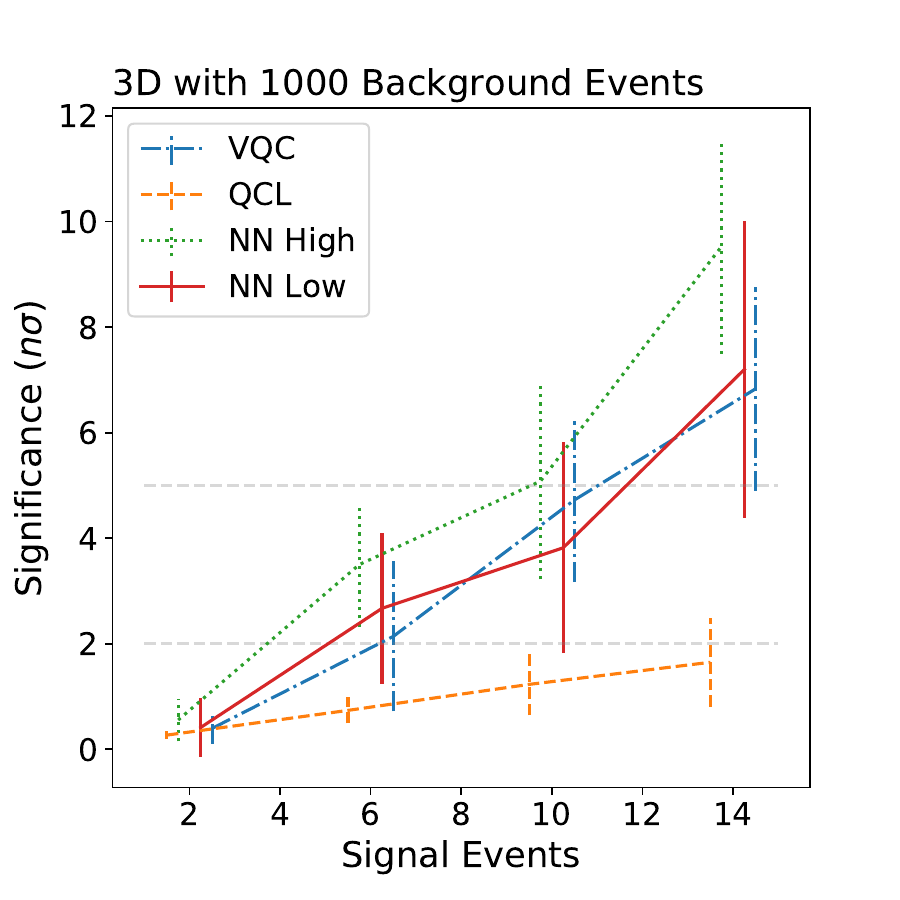}\includegraphics[width=0.5\textwidth]{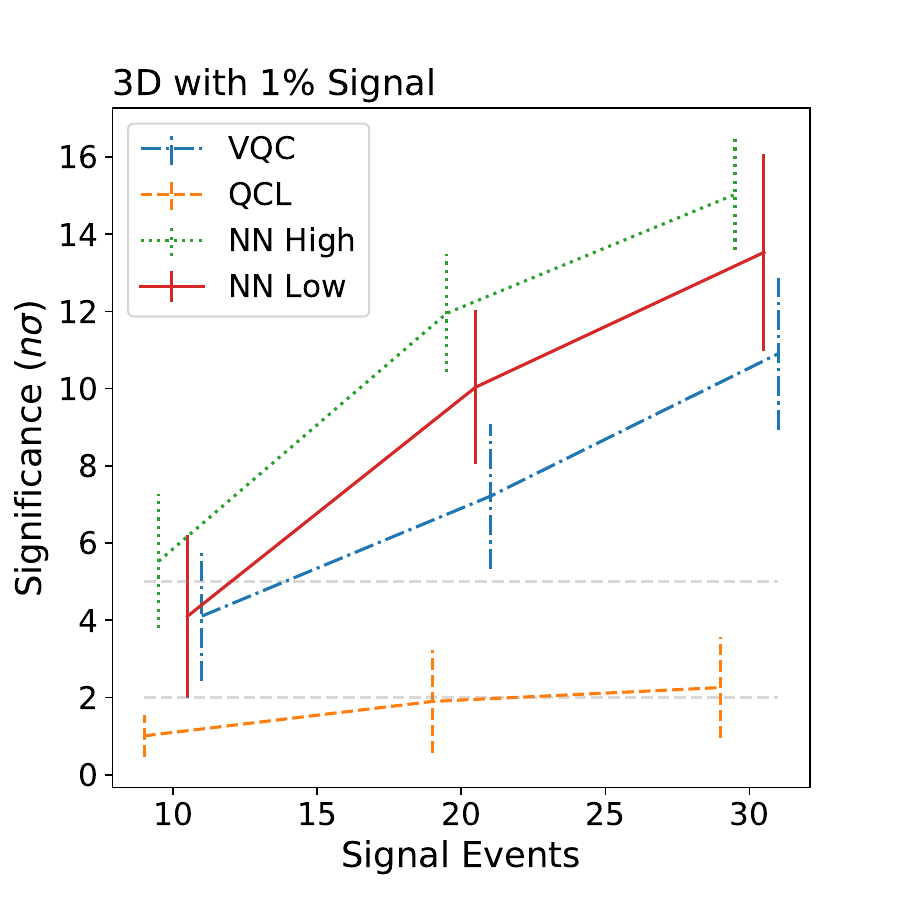}
    \caption{Results plotted against increasing signal events. The horizontal grey lines indicate thresholds for exclusion (2$\sigma$) and discovery (5$\sigma$). 
    }
    \label{fig:ratiowithfactor}
\end{figure}

Numerical results are presented in Fig.~\ref{fig:ratiowithfactor}.  As a first test, we fix the background at 1000 events and scan the number of signal events.  For reference, the naive significance with 30 signal events is about unity.  All of the methods are better than doing nothing and surpass $2\sigma$ by 40 events.  The QML models do not outperform the classical approaches and in fact appear to be systematically worse except perhaps for the lowest number of signal events where the significances themselves are below unity (and thus irrelevant).  Similar trends hold for the three-dimensional data, except that the QCL performs relatively worse and the NN High performs relatively better.  We note that the error bars on the plots are not small, which illustrates the importance of ensembling over the random parts of the training as well as over the random injetion of signal (and background) events.

As a second test, we fix the signal fraction and vary the number of signal events (which also changes the number of background events).  At 1\% fraction, for reference, 10 signal events would then correspond to the 1000 background events in the lefthand plots of Fig.~\ref{fig:ratiowithfactor}.  The trends for the fixed signal fraction are similar to the fixed background fraction, with slightly worse performance due to the larger number of background events beyond 10 signal events.  Additional performance metrics including the relative significance improvement and the Area Under the ROC curve can be found in Appendix~\ref{sec:appendix}.  These additional statistics corroborate the story presented in Fig.~\ref{fig:ratiowithfactor}.  The AUC is difficult to interpret as it integrates across the entire ROC curve when in practice, we typically operate at a fixed working point.  Small changes in the AUC could correspond to regions of the ROC curve that are physically useful, so the maximum significance improvement is instead chosen as the default statistic.

%%%%%%%%%%%%%%%%%%%%%%%%%%%%%%%%%%%%%%%%
\section{Conclusions}
\label{sec:concl}

Motivated by promising numerical studies from the HEP literature on quantum machine learning for low-event count applications, we have explored where these techniques could be practically useful for collider physics.  Given that most machine learning methods are trained using simulation, we concluded that a task relying directly on data should be the target application.  An important topic in this area is anomaly detection, where machine learning methods are used to reduce model dependence.  One strategy that reduces signal model dependence is to train a classifier to distinguish data from background-only simulation.  This will be fundamentally limited by the number of events in data and thus may be a good target for exploring an advantage of quantum machine learning over its classical counterpart.

There are not many final states that are modeled precisely enough for the data versus simulation strategy to be effective at finding small signals.  Following Ref.~\cite{Krzyzanska:2022mto}, we consider anomaly detection in the four-lepton final state, where the Standard Model is well-known, yet there is plenty of phase space for new physics.  We consider a low-dimensional version of the problem within a model framework that has three free parameters (masses), which already extends beyond the current searches at the LHC that focus on one and sometimes two-dimensional versions of the problem.  We do not find any advantage of quantum machine learning over classical machine learning.  

It could be that this particular problem is not well-suited for quantum machine learning or that we have not picked exactly the right quantum machine learning architecture or training process.  We do not claim that our results are general for all of QML and all of HEP, but we hope that our process and numerical results will be useful to put existing and future studies of QML for HEP in context.

%===================================================================
\section*{\label{sec::acknowledgments}Acknowledgments}
%===================================================================

This work was supported by the Department of Energy, Office of Science under contract number DE-AC02-05CH11231. S.A. was also supported by a University of California Summer Undergraduate Research Fellowship.  This research used resources of the Oak Ridge Leadership Computing Facility, which is a DOE Office of Science User Facility supported under Contract DE-AC05-00OR22725.  We thank Vinicius Mikuni for computing help and Andrew Blance, Koji Terashi, and Michael Spannowsky for useful discussions about QML.

%%%%%%%%%%%%%%%%%%%%%%%%%%%%%%%%%%%%%%%%
\section{Appendix}
\label{sec:appendix}

\begin{figure}[h!]
    \centering
    \includegraphics[width=0.5\textwidth]{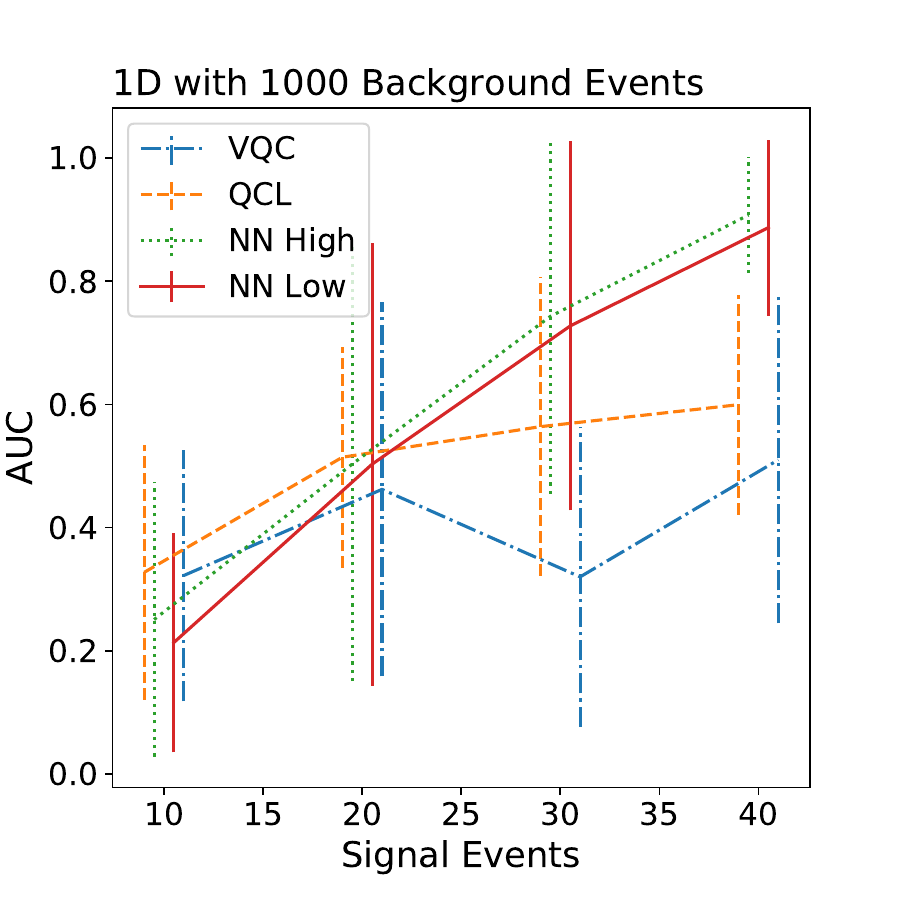}\includegraphics[width=0.5\textwidth]{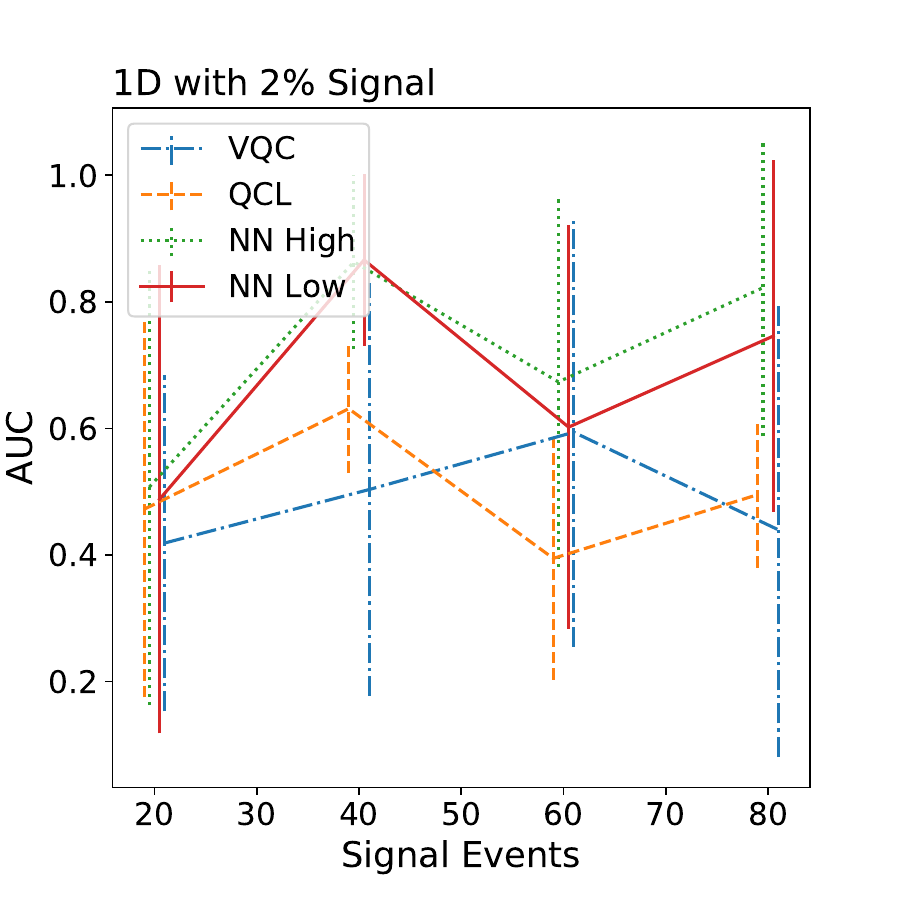}\\\includegraphics[width=0.5\textwidth]{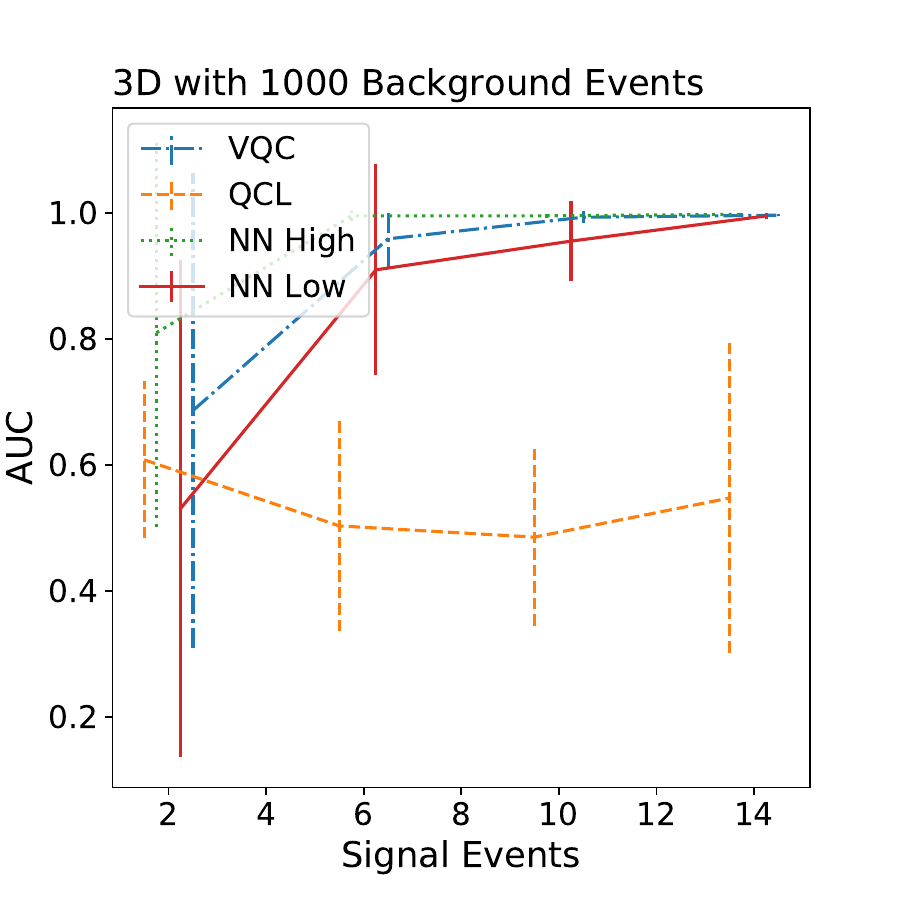}\includegraphics[width=0.5\textwidth]{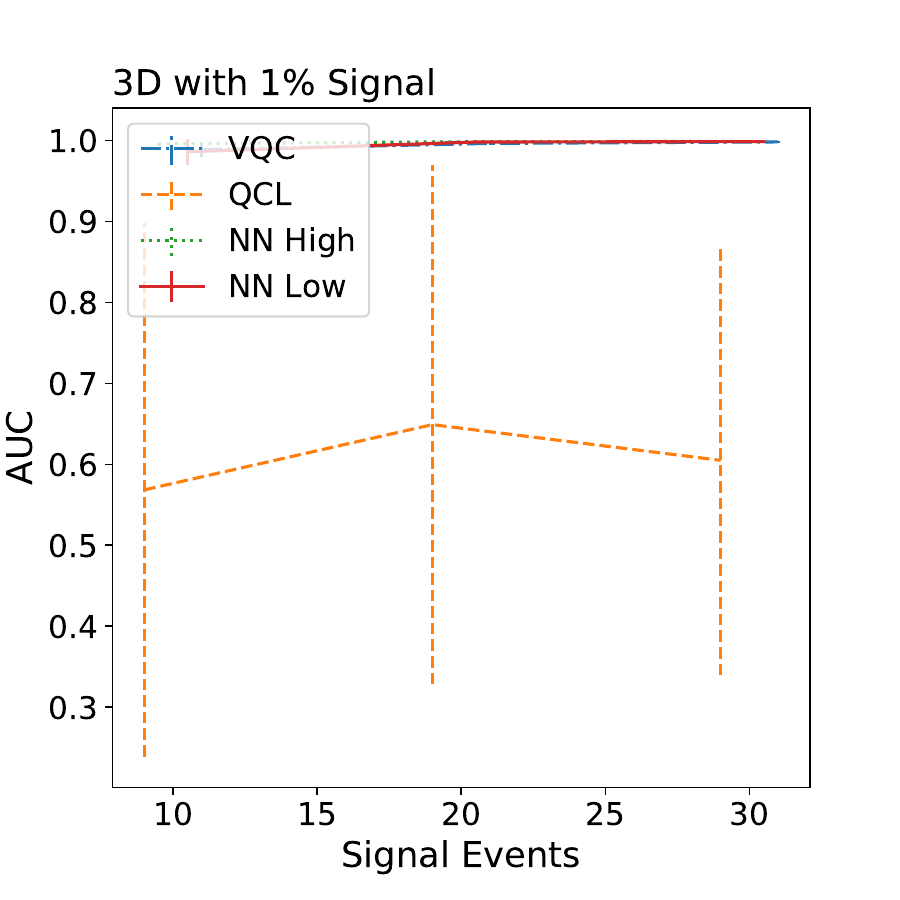}
    \caption{AUC plotted against increasing signal events}
    \label{fig:auc}
\end{figure}

\begin{figure}[h!]
    \centering
    \includegraphics[width=0.5\textwidth]{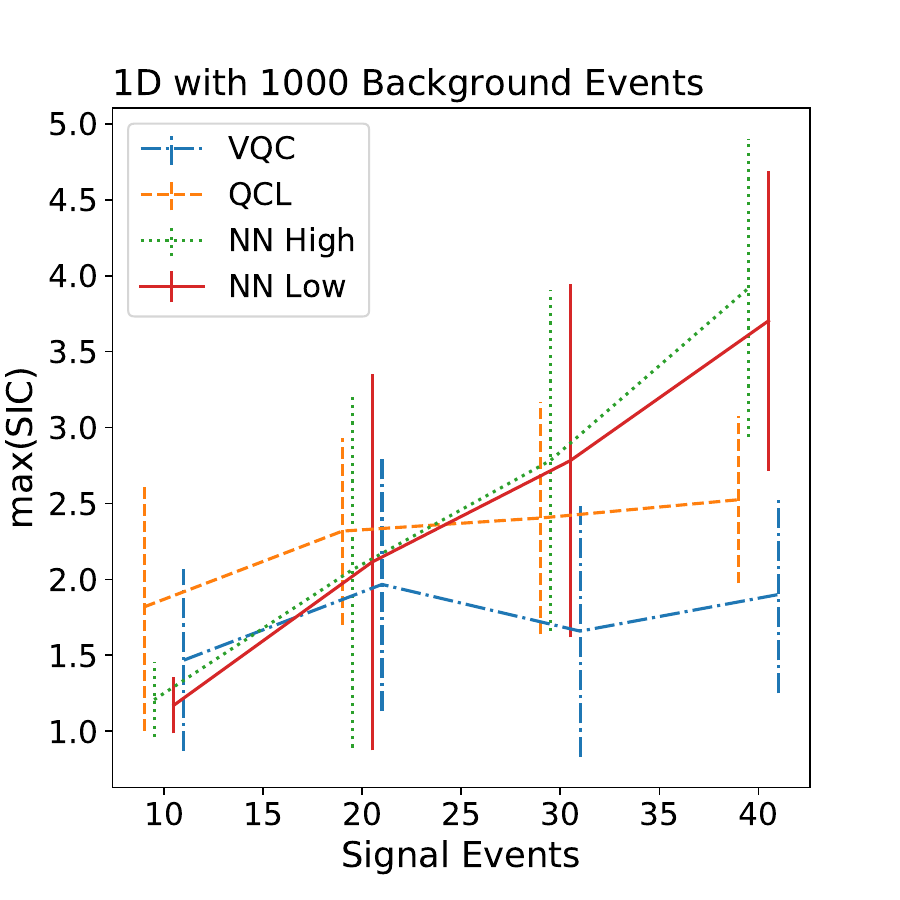}\includegraphics[width=0.5\textwidth]{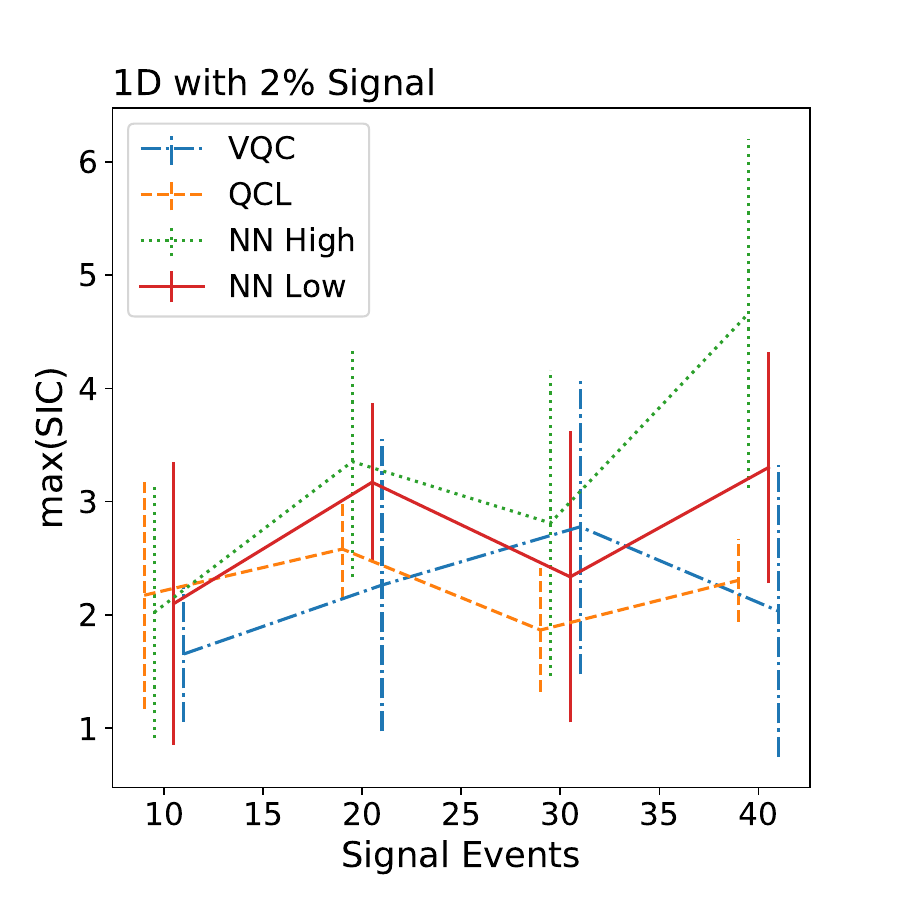}\\\includegraphics[width=0.5\textwidth]{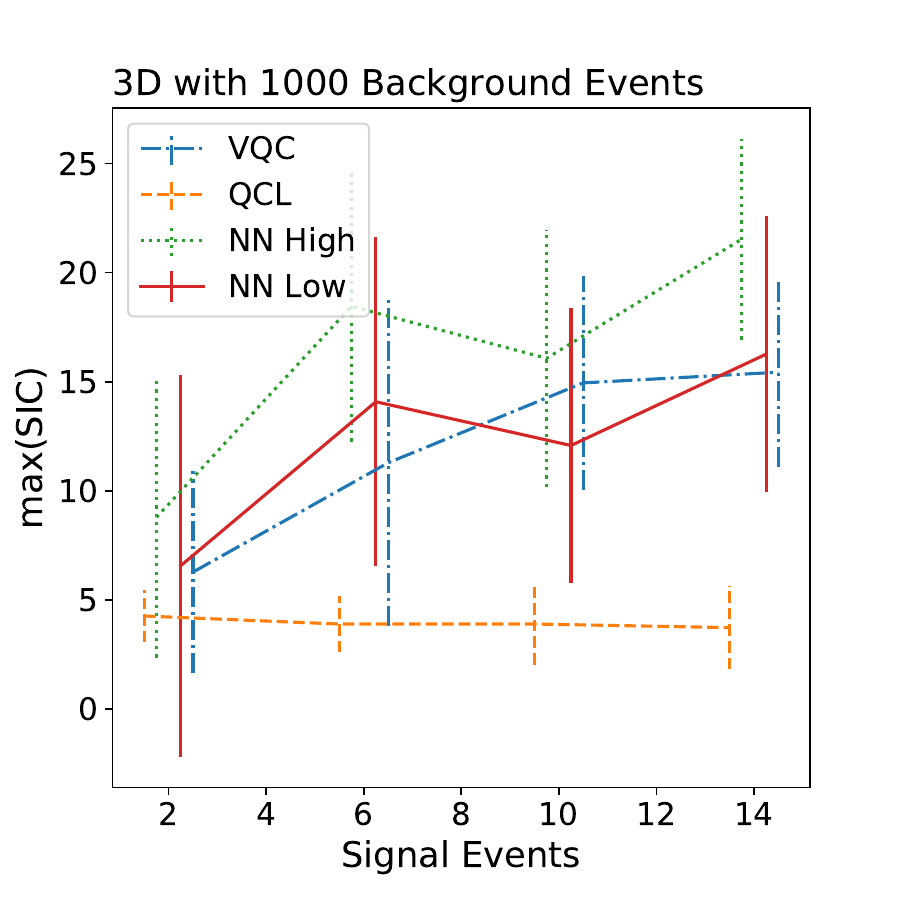}\includegraphics[width=0.5\textwidth]{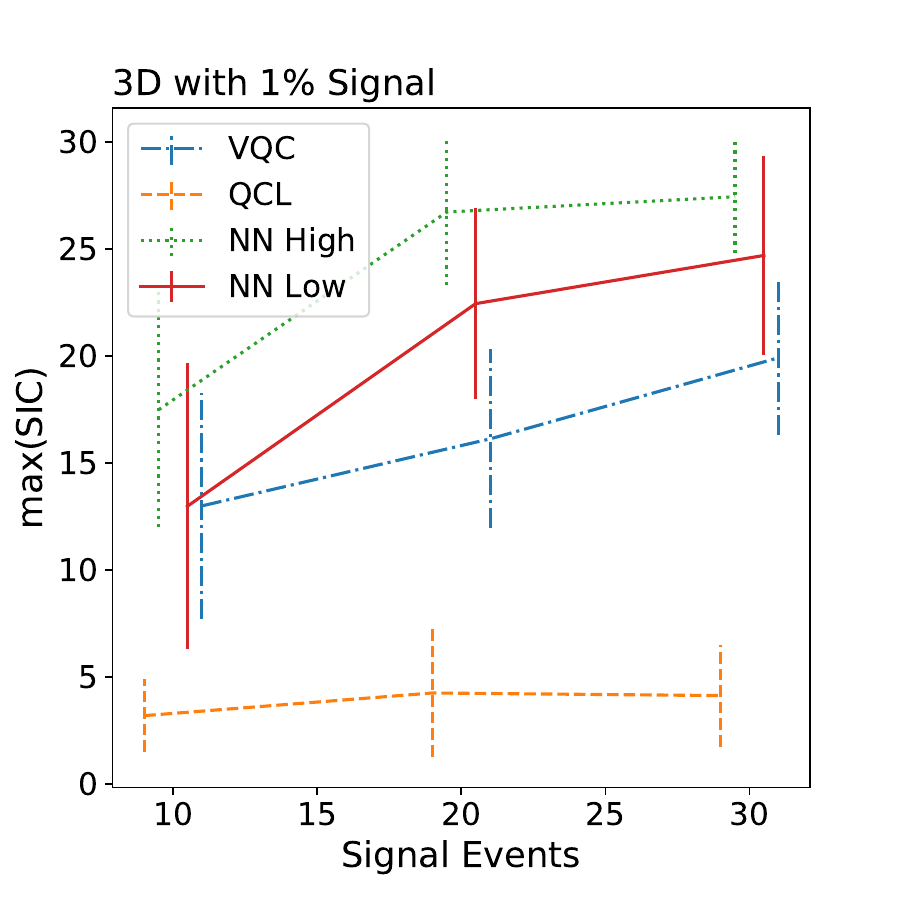}
    \caption{The maximum significance improvement as a function of the number of signal events.}
    \label{fig:rationofactor}
\end{figure}

Figures~\ref{fig:auc} and~\ref{fig:rationofactor} quantify the performance of the same studies as in Sec.~\ref{sec:mlsetup}, but using different metrics (Area Under the ROC curve and maximum significance improvement, respectively).

%%%%%%%%%%%%%%%%%%%%%%%%%%%%%%%%%%%%%%%%

\clearpage

\bibliographystyle{JHEP}
\bibliography{main,HEPML}

\providecommand{\href}[2]{#2}\begingroup\raggedright\begin{thebibliography}{100}

\bibitem{Jordan:2012xnu}
S.~P. Jordan, K.~S.~M. Lee and J.~Preskill, \emph{{Quantum Algorithms for
  Quantum Field Theories}},
  \href{https://doi.org/10.1126/science.1217069}{\emph{Science} {\bfseries 336}
  (2012) 1130--1133}, [\href{https://arxiv.org/abs/1111.3633}{{\ttfamily
  1111.3633}}].

\bibitem{Bauer:2022hpo}
C.~W. Bauer et~al., \emph{{Quantum Simulation for High Energy Physics}},
  \href{https://arxiv.org/abs/2204.03381}{{\ttfamily 2204.03381}}.

\bibitem{Mott:2017xdb}
A.~Mott, J.~Job, J.~R. Vlimant, D.~Lidar and M.~Spiropulu, \emph{{Solving a
  Higgs optimization problem with quantum annealing for machine learning}},
  \href{https://doi.org/10.1038/nature24047}{\emph{Nature} {\bfseries 550}
  (2017) 375--379}.

\bibitem{Zlokapa:2019lvv}
A.~Zlokapa, A.~Mott, J.~Job, J.-R. Vlimant, D.~Lidar and M.~Spiropulu,
  \emph{{Quantum adiabatic machine learning with zooming}},
  \href{https://doi.org/10.1103/PhysRevA.102.062405}{\emph{Phys. Rev. A}
  {\bfseries 102} (2020) 062405},
  [\href{https://arxiv.org/abs/1908.04480}{{\ttfamily 1908.04480}}].

\bibitem{Blance:2020nhl}
A.~Blance and M.~Spannowsky, \emph{{Quantum Machine Learning for Particle
  Physics using a Variational Quantum Classifier}},
  \href{https://doi.org/10.1007/JHEP02(2021)212}{\emph{Journal of High Energy
  Physics} (2020) }, [\href{https://arxiv.org/abs/2010.07335}{{\ttfamily
  2010.07335}}].

\bibitem{Terashi:2020wfi}
K.~Terashi, M.~Kaneda, T.~Kishimoto, M.~Saito, R.~Sawada and J.~Tanaka,
  \emph{{Event Classification with Quantum Machine Learning in High-Energy
  Physics}}, \href{https://doi.org/10.1007/s41781-020-00047-7}{\emph{Comput.
  Softw. Big Sci.} {\bfseries 5} (2021) 2},
  [\href{https://arxiv.org/abs/2002.09935}{{\ttfamily 2002.09935}}].

\bibitem{Chen:2020zkj}
S.~Y.-C. Chen, T.-C. Wei, C.~Zhang, H.~Yu and S.~Yoo, \emph{{Quantum
  convolutional neural networks for high energy physics data analysis}},
  \href{https://doi.org/10.1103/PhysRevResearch.4.013231}{\emph{Phys. Rev.
  Res.} {\bfseries 4} (2022) 013231},
  [\href{https://arxiv.org/abs/2012.12177}{{\ttfamily 2012.12177}}].

\bibitem{Wu:2020cye}
S.~L. Wu et~al., \emph{{Application of quantum machine learning using the
  quantum variational classifier method to high energy physics analysis at the
  LHC on IBM quantum computer simulator and hardware with 10 qubits}},
  \href{https://doi.org/10.1088/1361-6471/ac1391}{\emph{J. Phys. G} {\bfseries
  48} (2021) 125003}, [\href{https://arxiv.org/abs/2012.11560}{{\ttfamily
  2012.11560}}].

\bibitem{Chen:2021ouz}
S.~Y.-C. Chen, T.-C. Wei, C.~Zhang, H.~Yu and S.~Yoo, \emph{{Hybrid
  Quantum-Classical Graph Convolutional Network}},
  \href{https://arxiv.org/abs/2101.06189}{{\ttfamily 2101.06189}}.

\bibitem{Heredge:2021vww}
J.~Heredge, C.~Hill, L.~Hollenberg and M.~Sevior, \emph{{Quantum Support Vector
  Machines for Continuum Suppression in B Meson Decays}},
  \href{https://doi.org/10.1007/s41781-021-00075-x}{\emph{Comput. Softw. Big
  Sci.} {\bfseries 5} (2021) 27},
  [\href{https://arxiv.org/abs/2103.12257}{{\ttfamily 2103.12257}}].

\bibitem{Wu:2021xsj}
S.~L. Wu et~al., \emph{{Application of quantum machine learning using the
  quantum kernel algorithm on high energy physics analysis at the LHC}},
  \href{https://doi.org/10.1103/PhysRevResearch.3.033221}{\emph{Phys. Rev.
  Res.} {\bfseries 3} (2021) 033221},
  [\href{https://arxiv.org/abs/2104.05059}{{\ttfamily 2104.05059}}].

\bibitem{Belis:2021zqi}
V.~Belis, S.~Gonz\'alez-Castillo, C.~Reissel, S.~Vallecorsa, E.~F. Combarro,
  G.~Dissertori et~al., \emph{{Higgs analysis with quantum classifiers}},
  \href{https://doi.org/10.1051/epjconf/202125103070}{\emph{EPJ Web Conf.}
  {\bfseries 251} (2021) 03070},
  [\href{https://arxiv.org/abs/2104.07692}{{\ttfamily 2104.07692}}].

\bibitem{Araz:2021ifk}
J.~Y. Araz and M.~Spannowsky, \emph{{Quantum-inspired event reconstruction with
  Tensor Networks: Matrix Product States}},
  \href{https://doi.org/10.1007/JHEP08(2021)112}{\emph{JHEP} {\bfseries 08}
  (2021) 112}, [\href{https://arxiv.org/abs/2106.08334}{{\ttfamily
  2106.08334}}].

\bibitem{Bravo-Prieto:2021ehz}
C.~Bravo-Prieto, J.~Baglio, M.~C\`e, A.~Francis, D.~M. Grabowska and
  S.~Carrazza, \emph{{Style-based quantum generative adversarial networks for
  Monte Carlo events}},  \href{https://arxiv.org/abs/2110.06933}{{\ttfamily
  2110.06933}}.

\bibitem{Blance:2021gcs}
A.~Blance and M.~Spannowsky, \emph{{Unsupervised event classification with
  graphs on classical and photonic quantum computers}},
  \href{https://doi.org/10.1007/JHEP08(2021)170}{\emph{JHEP} {\bfseries 21}
  (2020) 170}, [\href{https://arxiv.org/abs/2103.03897}{{\ttfamily
  2103.03897}}].

\bibitem{Ngairangbam:2021yma}
V.~S. Ngairangbam, M.~Spannowsky and M.~Takeuchi, \emph{{Anomaly detection in
  high-energy physics using a quantum autoencoder}},
  \href{https://doi.org/10.1103/PhysRevD.105.095004}{\emph{Phys. Rev. D}
  {\bfseries 105} (2022) 095004},
  [\href{https://arxiv.org/abs/2112.04958}{{\ttfamily 2112.04958}}].

\bibitem{Araz:2022haf}
J.~Y. Araz and M.~Spannowsky, \emph{{Classical versus Quantum: comparing Tensor
  Network-based Quantum Circuits on LHC data}},
  \href{https://arxiv.org/abs/2202.10471}{{\ttfamily 2202.10471}}.

\bibitem{Gianelle:2022unu}
A.~Gianelle, P.~Koppenburg, D.~Lucchesi, D.~Nicotra, E.~Rodrigues, L.~Sestini
  et~al., \emph{{Quantum Machine Learning for $b$-jet identification}},
  \href{https://arxiv.org/abs/2202.13943}{{\ttfamily 2202.13943}}.

\bibitem{Humble:2022vtm}
T.~S. Humble et~al., \emph{{Snowmass White Paper: Quantum Computing Systems and
  Software for High-energy Physics Research}},  in \emph{{2022 Snowmass Summer
  Study}}, 3, 2022, \href{https://arxiv.org/abs/2203.07091}{{\ttfamily
  2203.07091}}.

\bibitem{Guan:2020bdl}
W.~Guan, G.~Perdue, A.~Pesah, M.~Schuld, K.~Terashi, S.~Vallecorsa et~al.,
  \emph{{Quantum Machine Learning in High Energy Physics}},
  \href{https://arxiv.org/abs/2005.08582}{{\ttfamily 2005.08582}}.

\bibitem{Schuld:2022lss}
M.~Schuld and N.~Killoran, \emph{{Is quantum advantage the right goal for
  quantum machine learning?}},
  \href{https://arxiv.org/abs/2203.01340}{{\ttfamily 2203.01340}}.

\bibitem{Karagiorgi:2021ngt}
G.~Karagiorgi, G.~Kasieczka, S.~Kravitz, B.~Nachman and D.~Shih, \emph{{Machine
  Learning in the Search for New Fundamental Physics}},
  \href{https://doi.org/10.1038/s42254-022-00455-1}{\emph{Nat. Rev. Phys.}
  {\bfseries 4} (12, 2022) 399},
  [\href{https://arxiv.org/abs/2112.03769}{{\ttfamily 2112.03769}}].

\bibitem{Collins:2018epr}
J.~H. Collins, K.~Howe and B.~Nachman, \emph{{Anomaly Detection for Resonant
  New Physics with Machine Learning}},
  \href{https://doi.org/10.1103/PhysRevLett.121.241803}{\emph{Phys. Rev. Lett.}
  {\bfseries 121} (2018) 241803},
  [\href{https://arxiv.org/abs/1805.02664}{{\ttfamily 1805.02664}}].

\bibitem{Collins:2019jip}
J.~H. Collins, K.~Howe and B.~Nachman, \emph{{Extending the search for new
  resonances with machine learning}},
  \href{https://doi.org/10.1103/PhysRevD.99.014038}{\emph{Phys. Rev.}
  {\bfseries D99} (2019) 014038},
  [\href{https://arxiv.org/abs/1902.02634}{{\ttfamily 1902.02634}}].

\bibitem{DAgnolo:2018cun}
R.~T. D'Agnolo and A.~Wulzer, \emph{{Learning New Physics from a Machine}},
  \href{https://doi.org/10.1103/PhysRevD.99.015014}{\emph{Phys. Rev.}
  {\bfseries D99} (2019) 015014},
  [\href{https://arxiv.org/abs/1806.02350}{{\ttfamily 1806.02350}}].

\bibitem{DAgnolo:2019vbw}
R.~T. D'Agnolo, G.~Grosso, M.~Pierini, A.~Wulzer and M.~Zanetti,
  \emph{{Learning multivariate new physics}},
  \href{https://doi.org/10.1140/epjc/s10052-021-08853-y}{\emph{Eur. Phys. J. C}
  {\bfseries 81} (2021) 89},
  [\href{https://arxiv.org/abs/1912.12155}{{\ttfamily 1912.12155}}].

\bibitem{Andreassen:2020nkr}
A.~Andreassen, B.~Nachman and D.~Shih, \emph{{Simulation Assisted
  Likelihood-free Anomaly Detection}},
  \href{https://doi.org/10.1103/PhysRevD.101.095004}{\emph{Phys. Rev. D}
  {\bfseries 101} (2020) 095004},
  [\href{https://arxiv.org/abs/2001.05001}{{\ttfamily 2001.05001}}].

\bibitem{Nachman:2020lpy}
B.~Nachman and D.~Shih, \emph{{Anomaly Detection with Density Estimation}},
  \href{https://doi.org/10.1103/PhysRevD.101.075042}{\emph{Phys. Rev. D}
  {\bfseries 101} (2020) 075042},
  [\href{https://arxiv.org/abs/2001.04990}{{\ttfamily 2001.04990}}].

\bibitem{Hallin:2021wme}
A.~Hallin, J.~Isaacson, G.~Kasieczka, C.~Krause, B.~Nachman, T.~Quadfasel
  et~al., \emph{{Classifying Anomalies THrough Outer Density Estimation
  (CATHODE)}},  \href{https://arxiv.org/abs/2109.00546}{{\ttfamily
  2109.00546}}.

\bibitem{Farina:2018fyg}
M.~Farina, Y.~Nakai and D.~Shih, \emph{{Searching for New Physics with Deep
  Autoencoders}},
  \href{https://doi.org/10.1103/PhysRevD.101.075021}{\emph{Phys. Rev. D}
  {\bfseries 101} (2020) 075021},
  [\href{https://arxiv.org/abs/1808.08992}{{\ttfamily 1808.08992}}].

\bibitem{Heimel:2018mkt}
T.~Heimel, G.~Kasieczka, T.~Plehn and J.~M. Thompson, \emph{{QCD or What?}},
  \href{https://doi.org/10.21468/SciPostPhys.6.3.030}{\emph{SciPost Phys.}
  {\bfseries 6} (2019) 030},
  [\href{https://arxiv.org/abs/1808.08979}{{\ttfamily 1808.08979}}].

\bibitem{Roy:2019jae}
T.~S. Roy and A.~H. Vijay, \emph{{A robust anomaly finder based on
  autoencoders}},  \href{https://arxiv.org/abs/1903.02032}{{\ttfamily
  1903.02032}}.

\bibitem{Cerri:2018anq}
O.~Cerri, T.~Q. Nguyen, M.~Pierini, M.~Spiropulu and J.-R. Vlimant,
  \emph{{Variational Autoencoders for New Physics Mining at the Large Hadron
  Collider}}, \href{https://doi.org/10.1007/JHEP05(2019)036}{\emph{JHEP}
  {\bfseries 05} (2019) 036},
  [\href{https://arxiv.org/abs/1811.10276}{{\ttfamily 1811.10276}}].

\bibitem{Blance:2019ibf}
A.~Blance, M.~Spannowsky and P.~Waite, \emph{{Adversarially-trained
  autoencoders for robust unsupervised new physics searches}},
  \href{https://doi.org/10.1007/JHEP10(2019)047}{\emph{JHEP} {\bfseries 10}
  (2019) 047}, [\href{https://arxiv.org/abs/1905.10384}{{\ttfamily
  1905.10384}}].

\bibitem{Hajer:2018kqm}
J.~Hajer, Y.-Y. Li, T.~Liu and H.~Wang, \emph{{Novelty Detection Meets Collider
  Physics}}, \href{https://doi.org/10.1103/PhysRevD.101.076015}{\emph{Phys.
  Rev. D} {\bfseries 101} (2020) 076015},
  [\href{https://arxiv.org/abs/1807.10261}{{\ttfamily 1807.10261}}].

\bibitem{DeSimone:2018efk}
A.~De~Simone and T.~Jacques, \emph{{Guiding New Physics Searches with
  Unsupervised Learning}},
  \href{https://doi.org/10.1140/epjc/s10052-019-6787-3}{\emph{Eur. Phys. J.}
  {\bfseries C79} (2019) 289},
  [\href{https://arxiv.org/abs/1807.06038}{{\ttfamily 1807.06038}}].

\bibitem{Mullin:2019mmh}
A.~Mullin, S.~Nicholls, H.~Pacey, M.~Parker, M.~White and S.~Williams,
  \emph{{Does SUSY have friends? A new approach for LHC event analysis}},
  \href{https://doi.org/10.1007/JHEP02(2021)160}{\emph{JHEP} {\bfseries 02}
  (2021) 160}, [\href{https://arxiv.org/abs/1912.10625}{{\ttfamily
  1912.10625}}].

\bibitem{1809.02977}
G.~M. Alessandro~Casa, \emph{{Nonparametric semisupervised classification for
  signal detection in high energy physics}},
  \href{https://arxiv.org/abs/1809.02977}{{\ttfamily 1809.02977}}.

\bibitem{Dillon:2019cqt}
B.~M. Dillon, D.~A. Faroughy and J.~F. Kamenik, \emph{{Uncovering latent jet
  substructure}},
  \href{https://doi.org/10.1103/PhysRevD.100.056002}{\emph{Phys. Rev.}
  {\bfseries D100} (2019) 056002},
  [\href{https://arxiv.org/abs/1904.04200}{{\ttfamily 1904.04200}}].

\bibitem{Aguilar-Saavedra:2017rzt}
J.~A. Aguilar-Saavedra, J.~H. Collins and R.~K. Mishra, \emph{{A generic
  anti-QCD jet tagger}},
  \href{https://doi.org/10.1007/JHEP11(2017)163}{\emph{JHEP} {\bfseries 11}
  (2017) 163}, [\href{https://arxiv.org/abs/1709.01087}{{\ttfamily
  1709.01087}}].

\bibitem{Romao:2019dvs}
M.~Romão~Crispim, N.~Castro, R.~Pedro and T.~Vale, \emph{{Transferability of
  Deep Learning Models in Searches for New Physics at Colliders}},
  \href{https://doi.org/10.1103/PhysRevD.101.035042}{\emph{Phys.\ Rev.\ D}
  {\bfseries 101} (2020) 035042},
  [\href{https://arxiv.org/abs/1912.04220}{{\ttfamily 1912.04220}}].

\bibitem{Romao:2020ojy}
M.~Crispim~Romão, N.~F. Castro, J.~G. Milhano, R.~Pedro and T.~Vale,
  \emph{{Use of a generalized energy Mover's distance in the search for rare
  phenomena at colliders}},
  \href{https://doi.org/10.1140/epjc/s10052-021-08891-6}{\emph{Eur. Phys. J. C}
  {\bfseries 81} (2021) 192},
  [\href{https://arxiv.org/abs/2004.09360}{{\ttfamily 2004.09360}}].

\bibitem{knapp2020adversarially}
O.~Knapp, O.~Cerri, G.~Dissertori, T.~Q. Nguyen, M.~Pierini and J.-R. Vlimant,
  \emph{{Adversarially Learned Anomaly Detection on CMS Open Data:
  re-discovering the top quark}},
  \href{https://doi.org/10.1140/epjp/s13360-021-01109-4}{\emph{Eur. Phys. J.
  Plus} {\bfseries 136} (2021) 236},
  [\href{https://arxiv.org/abs/2005.01598}{{\ttfamily 2005.01598}}].

\bibitem{collaboration2020dijet}
{\scshape ATLAS} collaboration, G.~Aad et~al., \emph{{Dijet resonance search
  with weak supervision using $\sqrt{s}=13$ TeV $pp$ collisions in the ATLAS
  detector}}, \href{https://doi.org/10.1103/PhysRevLett.125.131801}{\emph{Phys.
  Rev. Lett.} {\bfseries 125} (2020) 131801},
  [\href{https://arxiv.org/abs/2005.02983}{{\ttfamily 2005.02983}}].

\bibitem{1797846}
B.~M. Dillon, D.~A. Faroughy, J.~F. Kamenik and M.~Szewc, \emph{{Learning the
  latent structure of collider events}},
  \href{https://doi.org/10.1007/JHEP10(2020)206}{\emph{JHEP} {\bfseries 10}
  (2020) 206}, [\href{https://arxiv.org/abs/2005.12319}{{\ttfamily
  2005.12319}}].

\bibitem{1800445}
M.~Crispim Rom\~ao, N.~F. Castro and R.~Pedro, \emph{{Finding New Physics
  without learning about it: Anomaly Detection as a tool for Searches at
  Colliders}},
  \href{https://doi.org/10.1140/epjc/s10052-021-09813-2}{\emph{Eur. Phys. J. C}
  {\bfseries 81} (2021) 27},
  [\href{https://arxiv.org/abs/2006.05432}{{\ttfamily 2006.05432}}].

\bibitem{Amram:2020ykb}
O.~Amram and C.~M. Suarez, \emph{{Tag N\textquoteright{} Train: a technique to
  train improved classifiers on unlabeled data}},
  \href{https://doi.org/10.1007/JHEP01(2021)153}{\emph{JHEP} {\bfseries 01}
  (2021) 153}, [\href{https://arxiv.org/abs/2002.12376}{{\ttfamily
  2002.12376}}].

\bibitem{Cheng:2020dal}
T.~Cheng, J.-F. Arguin, J.~Leissner-Martin, J.~Pilette and T.~Golling,
  \emph{{Variational Autoencoders for Anomalous Jet Tagging}},
  \href{https://arxiv.org/abs/2007.01850}{{\ttfamily 2007.01850}}.

\bibitem{Khosa:2020qrz}
C.~K. Khosa and V.~Sanz, \emph{{Anomaly Awareness}},
  \href{https://arxiv.org/abs/2007.14462}{{\ttfamily 2007.14462}}.

\bibitem{Thaprasop:2020mzp}
P.~Thaprasop, K.~Zhou, J.~Steinheimer and C.~Herold, \emph{{Unsupervised
  Outlier Detection in Heavy-Ion Collisions}},
  \href{https://doi.org/10.1088/1402-4896/abf214}{\emph{Phys. Scripta}
  {\bfseries 96} (2021) 064003},
  [\href{https://arxiv.org/abs/2007.15830}{{\ttfamily 2007.15830}}].

\bibitem{Alexander:2020mbx}
S.~Alexander, S.~Gleyzer, H.~Parul, P.~Reddy, M.~W. Toomey, E.~Usai et~al.,
  \emph{{Decoding Dark Matter Substructure without Supervision}},
  \href{https://arxiv.org/abs/2008.12731}{{\ttfamily 2008.12731}}.

\bibitem{aguilarsaavedra2020mass}
J.~A. Aguilar-Saavedra, F.~R. Joaquim and J.~F. Seabra, \emph{{Mass Unspecific
  Supervised Tagging (MUST) for boosted jets}},
  \href{https://doi.org/10.1007/JHEP03(2021)012}{\emph{JHEP} {\bfseries 03}
  (2021) 012}, [\href{https://arxiv.org/abs/2008.12792}{{\ttfamily
  2008.12792}}].

\bibitem{Aguilar-Saavedra:2021utu}
J.~A. Aguilar-Saavedra, \emph{{Anomaly detection from mass unspecific jet
  tagging}}, \href{https://doi.org/10.1140/epjc/s10052-022-10058-w}{\emph{Eur.
  Phys. J. C} {\bfseries 82} (2022) 130},
  [\href{https://arxiv.org/abs/2111.02647}{{\ttfamily 2111.02647}}].

\bibitem{1815227}
K.~Benkendorfer, L.~L. Pottier and B.~Nachman, \emph{{Simulation-assisted
  decorrelation for resonant anomaly detection}},
  \href{https://doi.org/10.1103/PhysRevD.104.035003}{\emph{Phys. Rev. D}
  {\bfseries 104} (2021) 035003},
  [\href{https://arxiv.org/abs/2009.02205}{{\ttfamily 2009.02205}}].

\bibitem{pol2020anomaly}
{Adrian Alan Pol and Victor Berger and Gianluca Cerminara and Cecile Germain
  and Maurizio Pierini}, \emph{{Anomaly Detection With Conditional Variational
  Autoencoders}},  \href{https://arxiv.org/abs/2010.05531}{{\ttfamily
  2010.05531}}.

\bibitem{Mikuni:2020qds}
V.~Mikuni and F.~Canelli, \emph{{Unsupervised clustering for collider
  physics}}, \href{https://doi.org/10.1103/PhysRevD.103.092007}{\emph{Phys.
  Rev. D} {\bfseries 103} (2021) 092007},
  [\href{https://arxiv.org/abs/2010.07106}{{\ttfamily 2010.07106}}].

\bibitem{vanBeekveld:2020txa}
M.~van Beekveld, S.~Caron, L.~Hendriks, P.~Jackson, A.~Leinweber, S.~Otten
  et~al., \emph{{Combining outlier analysis algorithms to identify new physics
  at the LHC}}, \href{https://doi.org/10.1007/JHEP09(2021)024}{\emph{JHEP}
  {\bfseries 09} (2021) 024},
  [\href{https://arxiv.org/abs/2010.07940}{{\ttfamily 2010.07940}}].

\bibitem{Park:2020pak}
S.~E. Park, D.~Rankin, S.-M. Udrescu, M.~Yunus and P.~Harris, \emph{{Quasi
  Anomalous Knowledge: Searching for new physics with embedded knowledge}},
  \href{https://doi.org/10.1007/JHEP06(2021)030}{\emph{JHEP} {\bfseries 21}
  (2020) 030}, [\href{https://arxiv.org/abs/2011.03550}{{\ttfamily
  2011.03550}}].

\bibitem{Faroughy:2020gas}
D.~A. Faroughy, \emph{{Uncovering hidden new physics patterns in collider
  events using Bayesian probabilistic models}},
  \href{https://doi.org/10.22323/1.390.0238}{\emph{PoS} {\bfseries ICHEP2020}
  (2021) 238}, [\href{https://arxiv.org/abs/2012.08579}{{\ttfamily
  2012.08579}}].

\bibitem{Stein:2020rou}
G.~Stein, U.~Seljak and B.~Dai, \emph{{Unsupervised in-distribution anomaly
  detection of new physics through conditional density estimation}},
  \href{https://arxiv.org/abs/2012.11638}{{\ttfamily 2012.11638}}.

\bibitem{Kasieczka:2021xcg}
G.~Kasieczka et~al., \emph{{The LHC Olympics 2020 a community challenge for
  anomaly detection in high energy physics}},
  \href{https://doi.org/10.1088/1361-6633/ac36b9}{\emph{Rept. Prog. Phys.}
  {\bfseries 84} (2021) 124201},
  [\href{https://arxiv.org/abs/2101.08320}{{\ttfamily 2101.08320}}].

\bibitem{Batson:2021agz}
J.~Batson, C.~G. Haaf, Y.~Kahn and D.~A. Roberts, \emph{{Topological
  Obstructions to Autoencoding}},
  \href{https://doi.org/10.1007/JHEP04(2021)280}{\emph{JHEP} {\bfseries 04}
  (2021) 280}, [\href{https://arxiv.org/abs/2102.08380}{{\ttfamily
  2102.08380}}].

\bibitem{Bortolato:2021zic}
B.~Bortolato, B.~M. Dillon, J.~F. Kamenik and A.~Smolkovi\v{c}, \emph{{Bump
  hunting in latent space}},
  \href{https://doi.org/10.1103/PhysRevD.105.115009}{\emph{Phys. Rev. D}
  {\bfseries 105} (2022) 115009},
  [\href{https://arxiv.org/abs/2103.06595}{{\ttfamily 2103.06595}}].

\bibitem{Collins:2021nxn}
J.~H. Collins, P.~Mart\'\i{}n-Ramiro, B.~Nachman and D.~Shih, \emph{{Comparing
  weak- and unsupervised methods for resonant anomaly detection}},
  \href{https://doi.org/10.1140/epjc/s10052-021-09389-x}{\emph{Eur. Phys. J. C}
  {\bfseries 81} (2021) 617},
  [\href{https://arxiv.org/abs/2104.02092}{{\ttfamily 2104.02092}}].

\bibitem{Dillon:2021nxw}
B.~M. Dillon, T.~Plehn, C.~Sauer and P.~Sorrenson, \emph{{Better Latent Spaces
  for Better Autoencoders}},
  \href{https://doi.org/10.21468/SciPostPhys.11.3.061}{\emph{SciPost Phys.}
  {\bfseries 11} (2021) 061},
  [\href{https://arxiv.org/abs/2104.08291}{{\ttfamily 2104.08291}}].

\bibitem{Finke:2021sdf}
T.~Finke, M.~Kr\"amer, A.~Morandini, A.~M\"uck and I.~Oleksiyuk,
  \emph{{Autoencoders for unsupervised anomaly detection in high energy
  physics}}, \href{https://doi.org/10.1007/JHEP06(2021)161}{\emph{JHEP}
  {\bfseries 06} (2021) 161},
  [\href{https://arxiv.org/abs/2104.09051}{{\ttfamily 2104.09051}}].

\bibitem{Shih:2021kbt}
D.~Shih, M.~R. Buckley, L.~Necib and J.~Tamanas, \emph{{via machinae: Searching
  for stellar streams using unsupervised machine learning}},
  \href{https://doi.org/10.1093/mnras/stab3372}{\emph{Mon. Not. Roy. Astron.
  Soc.} {\bfseries 509} (2021) 5992--6007},
  [\href{https://arxiv.org/abs/2104.12789}{{\ttfamily 2104.12789}}].

\bibitem{Atkinson:2021nlt}
O.~Atkinson, A.~Bhardwaj, C.~Englert, V.~S. Ngairangbam and M.~Spannowsky,
  \emph{{Anomaly detection with convolutional Graph Neural Networks}},
  \href{https://doi.org/10.1007/JHEP08(2021)080}{\emph{JHEP} {\bfseries 08}
  (2021) 080}, [\href{https://arxiv.org/abs/2105.07988}{{\ttfamily
  2105.07988}}].

\bibitem{Kahn:2021drv}
A.~Kahn, J.~Gonski, I.~Ochoa, D.~Williams and G.~Brooijmans, \emph{{Anomalous
  jet identification via sequence modeling}},
  \href{https://doi.org/10.1088/1748-0221/16/08/P08012}{\emph{JINST} {\bfseries
  16} (2021) P08012}, [\href{https://arxiv.org/abs/2105.09274}{{\ttfamily
  2105.09274}}].

\bibitem{Aarrestad:2021oeb}
T.~Aarrestad et~al., \emph{{The Dark Machines Anomaly Score Challenge:
  Benchmark Data and Model Independent Event Classification for the Large
  Hadron Collider}},
  \href{https://doi.org/10.21468/SciPostPhys.12.1.043}{\emph{SciPost Phys.}
  {\bfseries 12} (2022) 043},
  [\href{https://arxiv.org/abs/2105.14027}{{\ttfamily 2105.14027}}].

\bibitem{Chakravarti:2021svb}
P.~Chakravarti, M.~Kuusela, J.~Lei and L.~Wasserman, \emph{{Model-Independent
  Detection of New Physics Signals Using Interpretable Semi-Supervised
  Classifier Tests}},  \href{https://arxiv.org/abs/2102.07679}{{\ttfamily
  2102.07679}}.

\bibitem{Dorigo:2021iyy}
T.~Dorigo, M.~Fumanelli, C.~Maccani, M.~Mojsovska, G.~C. Strong and B.~Scarpa,
  \emph{{RanBox: Anomaly Detection in the Copula Space}},
  \href{https://arxiv.org/abs/2106.05747}{{\ttfamily 2106.05747}}.

\bibitem{Caron:2021wmq}
S.~Caron, L.~Hendriks and R.~Verheyen, \emph{{Rare and Different: Anomaly
  Scores from a combination of likelihood and out-of-distribution models to
  detect new physics at the LHC}},
  \href{https://doi.org/10.21468/SciPostPhys.12.2.077}{\emph{SciPost Phys.}
  {\bfseries 12} (2022) 077},
  [\href{https://arxiv.org/abs/2106.10164}{{\ttfamily 2106.10164}}].

\bibitem{Govorkova:2021hqu}
E.~Govorkova, E.~Puljak, T.~Aarrestad, M.~Pierini, K.~A. Wo\'zniak and
  J.~Ngadiuba, \emph{{LHC physics dataset for unsupervised New Physics
  detection at 40 MHz}},
  \href{https://doi.org/10.1038/s41597-022-01187-8}{\emph{Sci. Data} {\bfseries
  9} (2022) 118}, [\href{https://arxiv.org/abs/2107.02157}{{\ttfamily
  2107.02157}}].

\bibitem{Kasieczka:2021tew}
G.~Kasieczka, B.~Nachman and D.~Shih, \emph{{New Methods and Datasets for Group
  Anomaly Detection From Fundamental Physics}},  7, 2021,
  \href{https://arxiv.org/abs/2107.02821}{{\ttfamily 2107.02821}}.

\bibitem{dAgnolo:2021aun}
R.~T. d'Agnolo, G.~Grosso, M.~Pierini, A.~Wulzer and M.~Zanetti,
  \emph{{Learning new physics from an imperfect machine}},
  \href{https://doi.org/10.1140/epjc/s10052-022-10226-y}{\emph{Eur. Phys. J. C}
  {\bfseries 82} (2022) 275},
  [\href{https://arxiv.org/abs/2111.13633}{{\ttfamily 2111.13633}}].

\bibitem{Volkovich:2021txe}
S.~Volkovich, F.~De~Vito~Halevy and S.~Bressler, \emph{{A data-directed
  paradigm for BSM searches: the bump-hunting example}},
  \href{https://doi.org/10.1140/epjc/s10052-022-10215-1}{\emph{Eur. Phys. J. C}
  {\bfseries 82} (2022) 265},
  [\href{https://arxiv.org/abs/2107.11573}{{\ttfamily 2107.11573}}].

\bibitem{Govorkova:2021utb}
E.~Govorkova et~al., \emph{{Autoencoders on field-programmable gate arrays for
  real-time, unsupervised new physics detection at 40 MHz at the Large Hadron
  Collider}}, \href{https://doi.org/10.1038/s42256-022-00441-3}{\emph{Nature
  Mach. Intell.} {\bfseries 4} (2022) 154--161},
  [\href{https://arxiv.org/abs/2108.03986}{{\ttfamily 2108.03986}}].

\bibitem{Ostdiek:2021bem}
B.~Ostdiek, \emph{{Deep Set Auto Encoders for Anomaly Detection in Particle
  Physics}}, \href{https://doi.org/10.21468/SciPostPhys.12.1.045}{\emph{SciPost
  Phys.} {\bfseries 12} (2022) 045},
  [\href{https://arxiv.org/abs/2109.01695}{{\ttfamily 2109.01695}}].

\bibitem{Fraser:2021lxm}
K.~Fraser, S.~Homiller, R.~K. Mishra, B.~Ostdiek and M.~D. Schwartz,
  \emph{{Challenges for unsupervised anomaly detection in particle physics}},
  \href{https://doi.org/10.1007/JHEP03(2022)066}{\emph{JHEP} {\bfseries 03}
  (2022) 066}, [\href{https://arxiv.org/abs/2110.06948}{{\ttfamily
  2110.06948}}].

\bibitem{Raine:2022hht}
J.~A. Raine, S.~Klein, D.~Sengupta and T.~Golling, \emph{{CURTAINs for your
  Sliding Window: Constructing Unobserved Regions by Transforming Adjacent
  Intervals}},  \href{https://arxiv.org/abs/2203.09470}{{\ttfamily
  2203.09470}}.

\bibitem{Krzyzanska:2022mto}
K.~Krzy\.za\'nska and B.~Nachman, \emph{{Simulation-based Anomaly Detection for
  Multileptons at the LHC}},
  \href{https://arxiv.org/abs/2203.09601}{{\ttfamily 2203.09601}}.

\bibitem{Letizia:2022xbe}
M.~Letizia, G.~Losapio, M.~Rando, G.~Grosso, A.~Wulzer, M.~Pierini et~al.,
  \emph{{Learning new physics efficiently with nonparametric methods}},
  \href{https://arxiv.org/abs/2204.02317}{{\ttfamily 2204.02317}}.

\bibitem{Letizia2021}
M.~Letizia et~al., \emph{{Efficient kernel methods for model-independent new
  physics searches}},  2021,
  \href{{https://ml4physicalsciences.github.io/2021/files/NeurIPS\_ML4PS\_2021\_146.pdf}}{{https://ml4physicalsciences.github.io/2021/files/NeurIPS\_ML4PS\_2021\_146.pdf}}.

\bibitem{ATLAS:2020tlo}
{\scshape ATLAS} collaboration, G.~Aad et~al., \emph{{Search for heavy
  resonances decaying into a pair of Z bosons in the $\ell ^+\ell ^-\ell
  '^+\ell '^-$ and $\ell ^+\ell ^-\nu {{\bar{\nu }}}$ final states using 139
  $\mathrm {fb}^{-1}$ of proton\textendash{}proton collisions at $\sqrt{s} =
  13\,$TeV with the ATLAS detector}},
  \href{https://doi.org/10.1140/epjc/s10052-021-09013-y}{\emph{Eur. Phys. J. C}
  {\bfseries 81} (2021) 332},
  [\href{https://arxiv.org/abs/2009.14791}{{\ttfamily 2009.14791}}].

\bibitem{ATLAS:2018coo}
{\scshape ATLAS} collaboration, M.~Aaboud et~al., \emph{{Search for Higgs boson
  decays to beyond-the-Standard-Model light bosons in four-lepton events with
  the ATLAS detector at $\sqrt{s}=13$ TeV}},
  \href{https://doi.org/10.1007/JHEP06(2018)166}{\emph{JHEP} {\bfseries 06}
  (2018) 166}, [\href{https://arxiv.org/abs/1802.03388}{{\ttfamily
  1802.03388}}].

\bibitem{ATLAS:2020wny}
{\scshape ATLAS} collaboration, G.~Aad et~al., \emph{{Measurements of the Higgs
  boson inclusive and differential fiducial cross sections in the 4$\ell$ decay
  channel at $\sqrt{s}$ = 13 TeV}},
  \href{https://doi.org/10.1140/epjc/s10052-020-8223-0}{\emph{Eur. Phys. J. C}
  {\bfseries 80} (2020) 942},
  [\href{https://arxiv.org/abs/2004.03969}{{\ttfamily 2004.03969}}].

\bibitem{CMS:2016ilx}
{\scshape CMS} collaboration, A.~M. Sirunyan et~al., \emph{{Measurements of the
  Higgs boson width and anomalous $HVV$ couplings from on-shell and off-shell
  production in the four-lepton final state}},
  \href{https://doi.org/10.1103/PhysRevD.99.112003}{\emph{Phys. Rev. D}
  {\bfseries 99} (2019) 112003},
  [\href{https://arxiv.org/abs/1901.00174}{{\ttfamily 1901.00174}}].

\bibitem{CMS:2020bni}
{\scshape CMS} collaboration, A.~Tumasyan et~al., \emph{{Search for low-mass
  dilepton resonances in Higgs boson decays to four-lepton final states in
  proton\textendash{}proton collisions at $\sqrt{s}=13\,\text {TeV} $}},
  \href{https://doi.org/10.1140/epjc/s10052-022-10127-0}{\emph{Eur. Phys. J. C}
  {\bfseries 82} (2022) 290},
  [\href{https://arxiv.org/abs/2111.01299}{{\ttfamily 2111.01299}}].

\bibitem{CMS:2021nnc}
{\scshape CMS} collaboration, A.~M. Sirunyan et~al., \emph{{Constraints on
  anomalous Higgs boson couplings to vector bosons and fermions in its
  production and decay using the four-lepton final state}},
  \href{https://doi.org/10.1103/PhysRevD.104.052004}{\emph{Phys. Rev. D}
  {\bfseries 104} (2021) 052004},
  [\href{https://arxiv.org/abs/2104.12152}{{\ttfamily 2104.12152}}].

\bibitem{Rebentrost_2014}
P.~Rebentrost, M.~Mohseni and S.~Lloyd, \emph{Quantum support vector machine
  for big data classification},
  \href{https://doi.org/10.1103/physrevlett.113.130503}{\emph{Phys. Rev. Lett.}
  {\bfseries 113} (2014) }.

\bibitem{PhysRevLett.103.150502}
A.~W. Harrow, A.~Hassidim and S.~Lloyd, \emph{Quantum algorithm for linear
  systems of equations},
  \href{https://doi.org/10.1103/PhysRevLett.103.150502}{\emph{Phys. Rev. Lett.}
  {\bfseries 103} (2009) 150502}.

\bibitem{Biamonte_2017}
J.~Biamonte, P.~Wittek, N.~Pancotti, P.~Rebentrost, N.~Wiebe and S.~Lloyd,
  \emph{Quantum machine learning},
  \href{https://doi.org/10.1038/nature23474}{\emph{Nature} {\bfseries 549}
  (2017) 195--202}.

\bibitem{Paul-Aymeric}
P.-A. McRae and M.~Hilke, \emph{{Quantum-Enhanced Machine Learning for Covid-19
  and Anderson Insulator Predictions}},
  \href{https://arxiv.org/abs/2012.03472}{{\ttfamily 2012.03472}}.

\bibitem{Pennylane}
V.~Bergholm, J.~Izaac, M.~Schuld, C.~Gogolin, M.~S. Alam, S.~Ahmed et~al.,
  \emph{Pennylane: Automatic differentiation of hybrid quantum-classical
  computations},  \href{https://arxiv.org/abs/1811.04968}{{\ttfamily
  1811.04968}}.

\bibitem{Alwall:2014hca}
J.~Alwall, R.~Frederix, S.~Frixione, V.~Hirschi, F.~Maltoni, O.~Mattelaer
  et~al., \emph{{The automated computation of tree-level and next-to-leading
  order differential cross sections, and their matching to parton shower
  simulations}}, \href{https://doi.org/10.1007/JHEP07(2014)079}{\emph{JHEP}
  {\bfseries 07} (2014) 079},
  [\href{https://arxiv.org/abs/1405.0301}{{\ttfamily 1405.0301}}].

\bibitem{Sjostrand:2006za}
T.~Sjostrand, S.~Mrenna and P.~Z. Skands, \emph{{PYTHIA 6.4 Physics and
  Manual}}, \href{https://doi.org/10.1088/1126-6708/2006/05/026}{\emph{JHEP}
  {\bfseries 05} (2006) 026},
  [\href{https://arxiv.org/abs/hep-ph/0603175}{{\ttfamily hep-ph/0603175}}].

\bibitem{Sjostrand:2007gs}
T.~Sjostrand, S.~Mrenna and P.~Z. Skands, \emph{{A Brief Introduction to PYTHIA
  8.1}}, \href{https://doi.org/10.1016/j.cpc.2008.01.036}{\emph{Comput. Phys.
  Commun.} {\bfseries 178} (2008) 852--867},
  [\href{https://arxiv.org/abs/0710.3820}{{\ttfamily 0710.3820}}].

\bibitem{Sjostrand:2014zea}
T.~Sj\"ostrand, S.~Ask, J.~R. Christiansen, R.~Corke, N.~Desai, P.~Ilten
  et~al., \emph{{An introduction to PYTHIA 8.2}},
  \href{https://doi.org/10.1016/j.cpc.2015.01.024}{\emph{Comput. Phys. Commun.}
  {\bfseries 191} (2015) 159--177},
  [\href{https://arxiv.org/abs/1410.3012}{{\ttfamily 1410.3012}}].

\bibitem{Zyla:2020zbs}
{\scshape Particle Data Group} collaboration, P.~Zyla et~al., \emph{{Review of
  Particle Physics}}, \href{https://doi.org/10.1093/ptep/ptaa104}{\emph{PTEP}
  {\bfseries 2020} (2020) 083C01}.

\bibitem{Alwall:2006yp}
J.~Alwall et~al., \emph{{A Standard format for Les Houches event files}},
  \href{https://doi.org/10.1016/j.cpc.2006.11.010}{\emph{Comput. Phys. Commun.}
  {\bfseries 176} (2007) 300--304},
  [\href{https://arxiv.org/abs/hep-ph/0609017}{{\ttfamily hep-ph/0609017}}].

\bibitem{deFavereau:2013fsa}
{\scshape DELPHES 3} collaboration, J.~de~Favereau, C.~Delaere, P.~Demin,
  A.~Giammanco, V.~Lema{\^\i}tre, A.~Mertens et~al., \emph{{DELPHES 3, A
  modular framework for fast simulation of a generic collider experiment}},
  \href{https://doi.org/10.1007/JHEP02(2014)057}{\emph{JHEP} {\bfseries 02}
  (2014) 057}, [\href{https://arxiv.org/abs/1307.6346}{{\ttfamily 1307.6346}}].

\bibitem{Mertens:2015kba}
A.~Mertens, \emph{{New features in Delphes 3}},
  \href{https://doi.org/10.1088/1742-6596/608/1/012045}{\emph{J. Phys. Conf.
  Ser.} {\bfseries 608} (2015) 012045}.

\bibitem{Selvaggi:2014mya}
M.~Selvaggi, \emph{{DELPHES 3: A modular framework for fast-simulation of
  generic collider experiments}},
  \href{https://doi.org/10.1088/1742-6596/523/1/012033}{\emph{J. Phys. Conf.
  Ser.} {\bfseries 523} (2014) 012033}.

\bibitem{Robens:2019kga}
T.~Robens, T.~Stefaniak and J.~Wittbrodt, \emph{{Two-real-scalar-singlet
  extension of the SM: LHC phenomenology and benchmark scenarios}},
  \href{https://doi.org/10.1140/epjc/s10052-020-7655-x}{\emph{Eur. Phys. J. C}
  {\bfseries 80} (2020) 151},
  [\href{https://arxiv.org/abs/1908.08554}{{\ttfamily 1908.08554}}].

\bibitem{Powell1994}
M.~J.~D. Powell, \emph{A Direct Search Optimization Method That Models the
  Objective and Constraint Functions by Linear Interpolation}, pp.~51--67.
\newblock Springer Netherlands, Dordrecht, 1994.

\bibitem{tensorflow2015-whitepaper}
M.~A. et~al, \emph{{TensorFlow}: Large-scale machine learning on heterogeneous
  systems},  2015.

\bibitem{adam}
D.~P. Kingma and J.~Ba, \emph{Adam: A method for stochastic optimization},
  \href{https://arxiv.org/abs/1412.6980}{{\ttfamily 1412.6980}}.

\end{thebibliography}\endgroup

\end{document}